%% file: arxiv.tex
\documentclass[11pt]{article}

\usepackage[a4paper,margin=2.2cm]{geometry}
\usepackage[T1]{fontenc}
\usepackage[utf8]{inputenc}

\DeclareUnicodeCharacter{2013}{--}
\DeclareUnicodeCharacter{2014}{---}
\DeclareUnicodeCharacter{2018}{`}
\DeclareUnicodeCharacter{2019}{'}
\DeclareUnicodeCharacter{201C}{``}
\DeclareUnicodeCharacter{201D}{''}

\usepackage{microtype}
\usepackage{amsmath,amssymb,amsthm}
\usepackage{booktabs}
\usepackage{array}
\usepackage{multirow}
\usepackage{graphicx}
\usepackage{algorithm}
\usepackage{algpseudocode}
\usepackage{xcolor}
\usepackage[round,authoryear]{natbib}
\usepackage{caption}
\usepackage{etoolbox}
\usepackage{docmute}
\usepackage[colorlinks=true,allcolors=blue!55!black]{hyperref}

\graphicspath{{Fig/}{media/}}
\newcommand{\preprinttitle}{}
\newcommand{\preprintauthors}{}
\newcommand{\preprintaffiliations}{}
\newcommand{\preprintcorrespondence}{}
\newcommand{\preprintabstract}{}
\newcommand{\preprintkeywords}{}
\newcommand{\preprintgraphicalabstract}{}
\newif\ifpreprintfirstauthor
\preprintfirstauthortrue

\renewcommand{\title}[2][]{\gdef\preprinttitle{#2}}
\renewcommand{\author}[2][]{%
  \ifpreprintfirstauthor
    \gdef\preprintauthors{#2\textsuperscript{#1}}%
    \global\preprintfirstauthorfalse
  \else
    \gappto\preprintauthors{, #2\textsuperscript{#1}}%
  \fi
}
\newcommand{\address}[2][]{%
  \gappto\preprintaffiliations{\par\textsuperscript{#1}#2}%
}
\newcommand{\corresp}[2][]{\gdef\preprintcorrespondence{#2}}
\renewcommand{\abstract}[1]{\gdef\preprintabstract{#1}}
\newcommand{\keywords}[1]{\gdef\preprintkeywords{#1}}
\newcommand{\otherabstract}[2][]{\gdef\preprintgraphicalabstract{#2}}

\newcommand{\journaltitle}[1]{}
\newcommand{\DOI}[1]{}
\newcommand{\copyrightyear}[1]{}
\newcommand{\pubyear}[1]{}
\newcommand{\access}[1]{}
\newcommand{\appnotes}[1]{}
\newcommand{\firstpage}[1]{}
\newcommand{\received}[3]{}
\newcommand{\revised}[3]{}
\newcommand{\accepted}[3]{}
\newcommand{\orgdiv}[1]{#1}
\newcommand{\orgname}[1]{#1}
\newcommand{\orgaddress}[1]{#1}
\newcommand{\country}[1]{#1}

\renewcommand{\maketitle}{%
  \begin{center}
    {\LARGE\bfseries \preprinttitle\par}
    \vspace{0.9em}
    {\large \preprintauthors\par}
    \vspace{0.6em}
    {\small \preprintaffiliations\par}
    \ifstrempty{\preprintcorrespondence}{}{%
      \vspace{0.5em}{\small \preprintcorrespondence\par}%
    }
  \end{center}
  \vspace{0.8em}
  \begin{quote}
    \small
    \noindent\textbf{Abstract}\par\smallskip
    \preprintabstract
  \end{quote}
  \noindent\textbf{Keywords: }\preprintkeywords\par
  \ifstrempty{\preprintgraphicalabstract}{}{%
    \vspace{1em}
    \begin{center}
      \textbf{Graphical abstract}\par\medskip
      \preprintgraphicalabstract
    \end{center}
  }
  \vspace{1em}
}

\theoremstyle{plain}

\theoremstyle{definition}
\newtheorem{definition}{Definition}
\theoremstyle{remark}
\newtheorem{remark}{Remark}

\providecommand{\mathbbm}[1]{\mathbf{#1}}

\newcommand{\mat}[1]{\mathbf{#1}}
\newcommand{\ten}[1]{\boldsymbol{\mathcal{#1}}}
\newcommand{\supra}{\mat{A}}
\newcommand{\R}{\mathbb{R}}
\newcommand{\code}[1]{\texttt{#1}}

\let\preprintoriginalbibliography\bibliography
\let\preprintoriginalbibliographystyle\bibliographystyle
\renewcommand{\bibliography}[1]{\gdef\preprintbibdatabase{#1}}
\renewcommand{\bibliographystyle}[1]{\gdef\preprintbibstyle{#1}}

\begin{document}

%% docmute discards the nested document class/preamble and keeps the body.
%% main.tex therefore remains the manuscript source of truth.
\let\preprintoriginaladdcontentsline\addcontentsline
\renewcommand{\addcontentsline}[3]{}

\input{main.tex}

\let\addcontentsline\preprintoriginaladdcontentsline

%% Supplementary Information follows in the same arXiv document.
\clearpage
\setcounter{secnumdepth}{3}
\setcounter{section}{0}
\setcounter{subsection}{0}
\setcounter{figure}{0}
\setcounter{table}{0}
\setcounter{algorithm}{0}
\setcounter{definition}{0}
\setcounter{remark}{0}
\setcounter{equation}{0}
\renewcommand{\thesection}{S\arabic{section}}
\renewcommand{\thesubsection}{S\arabic{section}.\arabic{subsection}}
\renewcommand{\thefigure}{S\arabic{figure}}
\renewcommand{\thetable}{S\arabic{table}}
\renewcommand{\thealgorithm}{S\arabic{algorithm}}
\renewcommand{\thedefinition}{S\arabic{definition}}
\renewcommand{\theremark}{S\arabic{remark}}
\renewcommand{\theequation}{S\arabic{equation}}

\renewcommand{\maketitle}{%
  \begin{center}
    {\LARGE\bfseries Additional File 1 --- Supplementary Information\par}
    \vspace{0.5em}
    {\large muxvizpy: a Python library for the analysis of multilayer biological networks\par}
    \vspace{0.7em}
    {\normalsize Matteo Baldan, Francesco Zambelli, Andrea Valsecchi, Giulia Cesaro,\\
    Giacomo Baruzzo, Manlio De Domenico, Barbara Di Camillo\par}
  \end{center}
  \vspace{1em}
}

\input{supp.tex}

\clearpage
\let\bibliography\preprintoriginalbibliography
\let\bibliographystyle\preprintoriginalbibliographystyle
\bibliographystyle{oup-abbrvnat}
\bibliography{reference}

\end{document}

%% file: main.tex
\journaltitle{Bioinformatics}
\DOI{DOI added during production}
\copyrightyear{2026}
\pubyear{2026}
\access{Advance Access Publication Date: Day Month Year}
\appnotes{Application Note}

\firstpage{1}

\title[muxvizpy]{muxvizpy: a Python library for the analysis of multilayer biological networks}

%% TODO: fill in actual authors and affiliations
\author[1,$\dagger$,$\ast$]{Matteo Baldan}
\author[2,$\dagger$]{Francesco Zambelli}
\author[3]{Andrea Valsecchi}
\author[1]{Giacomo Baruzzo}
\author[1]{Giulia Cesaro}
\author[2,5]{Manlio De Domenico}
\author[1,4,5,$\ast$]{Barbara Di Camillo}

\address[1]{\orgdiv{Department of Information Engineering}, \orgname{University of Padova}, \orgaddress{\country{Italy}}}
\address[2]{\orgdiv{Department of Physics and Astronomy}, \orgname{University of Padova}, \orgaddress{\country{Padova}}}
\address[3]{\orgdiv{Department of Mathematics}, \orgname{University of Padova}, \orgaddress{\country{Padova}}}
\address[4]{\orgdiv{Department of Comparative Biomedicine and Food Science}, \orgname{University of Padova}, \orgaddress{\country{Padova}}}
\address[5]{\orgdiv{Padua Center for Network Medicine}, \orgname{University of Padova}, \orgaddress{\country{Padova}}}

\corresp[$\ast$]{Corresponding author: 
\href{mailto:barbara.dicamillo@unipd.it}{barbara.dicamillo@unipd.it}; $\dagger$: authors share contribution}
% Dates — fill in
\received{}{0}{}
\revised{}{0}{}
\accepted{}{0}{}

%% ----------------------------------------------------------------
%% ABSTRACT
%% ----------------------------------------------------------------
\abstract{%
Biological systems are inherently multilayered: the same entities---genes, cells, or bacterial species---participate simultaneously in qualitatively distinct types of interactions, each carrying complementary information that no single relational view can capture.
Analysing such systems with single-layer tools, or by collapsing layers into a monoplex projection, systematically discards inter-layer dependencies and can yield misleading conclusions about centrality, community structure, and robustness.
The multilayer network formalism addresses, and \texttt{muxViz} established one of the first comprehensive toolkits for its structural analysis, but its R-only interface and dense data structures limit applicability to large biological networks. We introduce \textit{muxvizpy}, a Python library that reimplements and extends the \texttt{muxViz} analytical catalogue with a sparse linear-algebra stack built on SciPy and PyTorch.
Muxvizpy exposes seven categories through a unified, composable API and is numerically validated against \texttt{muxViz} on synthetic Erd\H{o}s--R\'enyi and Barab\'asi--Albert multiplex networks while substantially reducing peak memory and wall-clock time at scale.
We illustrate its applicability on a virus--human protein-interaction multiplex in which computing some structural analysis was unfeasible.
\\[2pt]
muxvizpy is freely available under the MIT licence at
\url{https://github.com/CoMuNeLab/MuxVizPy}.
Mathematical definitions of all implemented metrics are provided in the Additional File.
}
\keywords{multilayer networks; network analysis; biological networks; complex systems; scientific computing}

\otherabstract[Graphical Abstract]{%
\centering
\includegraphics[width=\textwidth]{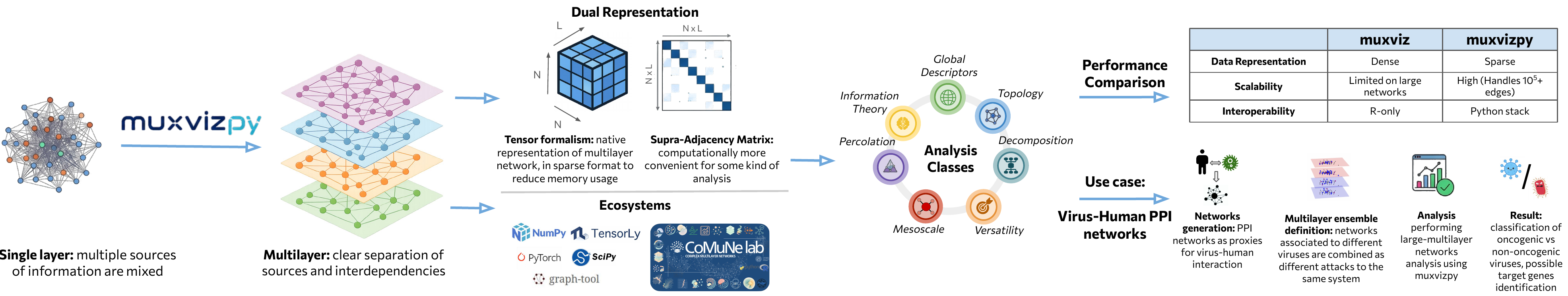}%
\par}

\maketitle

%% ================================================================
%% 1. INTRODUCTION
%% ================================================================
\section{Introduction}

% --- SECTION 1: From Foundation to Multilayer Necessity ---
The study of complex systems has evolved from analyzing isolated networks to a multilayer formalism that captures the interdependent interactions inherent in social, technological and biological systems~\citep{Kivela2014, DeDomenico2023, aleta2026multilayer}. The foundational work of Watts and Strogatz on small-world properties and Barabási and Albert on scale-free distributions established a robust mathematical basis for network science; however, it has since become clear that real-world entities rarely interact through single channels or in single contexts. The multilayer framework addresses this directly, representing a system as a collection of layers, each encoding a distinct mode of interaction among a set of entities (nodes) ~\citep{Kivela2014, DeDomenico2023}. The utility of this formalism has been demonstrated across a broad range of domains, from social, biological systems and power grids to multimodal transportation networks and communication infrastructure ~\citep{aleta2026multilayer}. 

% --- Paragraph 3: biological systems as a natural application domain ---
Biological systems are a particularly natural fit for this framework. At the cellular scale, the same gene can be involved in multiple biological processes and can be observed across qualitatively distinct interaction modalities that  naturally map to different layers: its expression captured by the transcriptomic layer, its regulatory state by the epigenomic layer, the physical interactions of its protein product by the protein--protein interaction (PPI) layer, and its metabolic activity by the metabolic layer \citep{aleta2026multilayer}. Each layer derives from a different omic technology and carries complementary biological information; their joint analysis can reveal disease modules, pathway cross-talk, and regulatory feedback loops that remain invisible when any single layer is examined in isolation. The same logic extends to larger scales, from synaptic and neuromodulatory coupling in neural circuits to trophic and competitive layers in ecosystems.

The increasing availability of high-throughput data and the growth of computational capacity have since elevated this framework from a theoretical convenience to a practical necessity for systems-level analysis at scale~\citep{aleta2026multilayer}. However, a significant software gap remains: while R-based tools like muxViz~\citep{DeDomenico2015muxViz} offer comprehensive metrics, they lack native interoperability with the dominant Python-based bioinformatics stack (e.g., Scanpy, PyTorch) and often fail to scale due to dense data representations~\citep{Kjellstrom2023comparative, Buth2025}.

% --- Paragraph 4: the software gap and motivation for\texttt{muxvizpy}---
The multilayer analytical toolkit is currently distributed across a fragmented ecosystem: in R, muxViz~\citep{DeDomenico2015muxViz} offers comprehensive structural metrics and 3D visualization, while multinet offers efficient C++ backed routines for actor-based measures and layer comparison.
In Python, Pymnet and multiNetX support general multilayer and multiplex structures and MultilayerGraphs.jl \cite{Kjellstrom2023comparative} extends the ecosystem to Julia. General-purpose libraries such as graph-tool ~\citep{peixoto2023descriptive} deliver excellent performance on monoplex graphs, but lack native multilayer data structures. Despite this breadth, no single tool covers the full analytical catalogue needed in bioinformatics while addressing two practical constraints: \emph{Python interoperability}, where the dominant pipelines for single-cell analysis (Scanpy/AnnData), graph deep learning (PyTorch Geometric), and multi-omics integration (MOFA+, MuData) are native, and \emph{scalability}, where existing multilayer libraries fail on large networks due to usage of dense representations ~\citep{Kjellstrom2023comparative}.

% --- Paragraph 5: what\texttt{muxvizpy}is and what this paper contributes ---
To address these constraints, we introduce \texttt{muxvizpy}, a Python library that not only re-implements the analytical catalogue of \texttt{muxViz} but also redesigns multilayer network analysis around sparse tensor algebra and native interoperability with the modern Python scientific ecosystem. Unlike existing multilayer software,\texttt{muxvizpy} combines (i) sparse-native representations for large-scale networks, (ii) exact and approximate solvers for computationally demanding centrality measures, (iii) tensor decomposition methods for multilayer structure discovery, and (iv) integration with tools such as SciPy, PyTorch, graph-tool and Scanpy. These features enable analyses that are computationally prohibitive in existing implementations while preserving compatibility with multilayer network methodologies. 
All metrics are numerically validated against \texttt{muxViz} on synthetic benchmarks, while its usage is highlighted by a biological use case related to the study of interactions between viruses and their human host with PPI networks.

%% ================================================================
%% 2. THE\texttt{muxvizpy}PACKAGE
%% ================================================================
\vspace{-8pt}
\section{The \texttt{muxvizpy} package}

\subsection{Design principles}

% --- core design paragraph ---
muxvizpy is built around three design principles that together address the limitations of existing tools.
\emph{Sparse-first.}
All supra-adjacency matrices and tensors are stored as SciPy CSR sparse matrices or PyTorch COO sparse tensors; inter-layer coupling blocks are either generated when not explicitly provided or read from edge lists. 
\emph{Composable.}
The library exposes a small set of fundamental data-structure constructors implemented in the parsing and i/o utilities, from which all higher-level metrics are derived; users can supply their own supra-adjacency matrix or use the mode-1 unfolding procedure to convert the tensor and call any metric directly.
\emph{Validated.}
Every metric ships with a numerical test against the corresponding \texttt{muxViz} output on a shared synthetic network, ensuring cross-platform reproducibility with some numerical precision threshold.

\vspace{-6pt}
\subsection{Data structures and structural metrics}

The central object in \texttt{muxvizpy} is the \emph{adjacency tensor} $\mathcal{M} \in \mathbb{R}^{N \times N \times L \times L}$, stored as a PyTorch COO sparse tensor, where $N$ is the number of nodes and $L$ the number of layers.
This representation is preferred over the supra-adjacency matrix because it preserves the dimensions of the network---namely $N$ and $L$---which are lost when the tensor is flattened into a single $(NL \times NL)$ matrix.
For the majority of structural metrics, the tensor is efficiently parsed via mode-1 unfolding into the \emph{supra-adjacency matrix} $\mathbf{A} \in \mathbb{R}^{NL \times NL}$, stored as a SciPy CSR sparse matrix, whose diagonal blocks of size $N \times N$ encode intra-layer adjacencies and whose off-diagonal blocks encode inter-layer couplings.
For spectral decomposition and tensor-native analyses, the adjacency tensor is used directly without unfolding. For node-aligned multiplex networks with categorical couplings (the most common case in practice), the coupling blocks reduce to scalar-weighted identity matrices, which \texttt{muxvizpy} stores implicitly.
Input is accepted as a five-column edge-list CSV (\texttt{node.from, layer.from, node.to, layer.to, weight}), compatible with the \texttt{muxViz} edge-list format.
The structural metrics implemented in \texttt{muxvizpy} are classified as versatility, mesoscale, topology, information theory, decomposition, global descriptors and percolation and are defined in the Extended table ~1. Mathematical definitions are provided in the Additional File~1.

\begin{paragraph}{Approximate against exact algorithms}
A central computational contribution of \texttt{muxvizpy} is the integration of alternative solution strategies for centrality measures whose evaluation becomes increasingly demanding as the number of node and layers grows. Many reduce either to solving a large sparse linear system or to identifying the dominant eigenvector of a supra-network operator. \texttt{muxvizpy} exposes exact and approximate computational paths within the same framework, allowing the numerical strategy to be selected according to network size, conditioning, and the desired balance between accuracy and computational cost \citep{NEURIPS2024_e88870ec,Saad2003iterative}.

For Katz and eigenvector centralities, \texttt{muxvizpy} provides an exact sparse-LU solution and two approximate alternatives. Approximate alternatives involve the Neumann-series method, yielding a matrix-free fixed-point iteration that converges below the spectral limit and a ILU-preconditioned GMRES and BiCGSTAB series that offer a scalable compromise, accelerating convergence while avoiding the memory cost of a complete LU factorisation.

PageRank and random-walk centralities are instead formulated as dominant-eigenvector problems for a stochastic transition operator, where the Perron-Frobenius theorem holds. Their stationary node--layer distribution can be obtained either with a sparse eigensolver or through power iteration. The contribution is therefore a unified computational framework that provides direct, matrix-free, and preconditioned iterative methods for the same multilayer centralities, extending their applicability to supra-adjacency matrices for which exact methods become limited by time or memory. Further details are provided in Additional File~1.
\end{paragraph}

% Several centrality measures in\texttt{muxvizpy}offer both exact and approximate computation paths with the hypothesis that for large network sizes many centralities are unfeasible and a power iteration approach, such as the Neumann series \citep{NEURIPS2024_e88870ec} and still provide accurate results. Katz centrality can be solved exactly via sparse LU factorisation or via iterative approximations like Neumann series to converge unconditionally or via Krylov iteration when preconditioned by an incomplete LU factorization. Random-walk centralities like PageRank centrality can be solved similarly considering that for row-stochastic transition matrices, the Perron-Frobenius theorem holds and we can find the centrality value of each node by solving the eigenvector decomposition of the transition matrix. Which, can be achieved again iteratively or by exact solution. Further details are found in Additional File 1.
% \end{paragraph}

% \paragraph{Integration with graph tools and the CoMuNe lab ecosystem}
%\texttt{muxvizpy}integrates tightly with graph-tool \citep{peixoto2023descriptive} to combine the strengths of a high-performance C++ graph library with a dedicated structural analysis framework for multilayer networks. This integration enables the computation of topological structures and mesoscale metrics such as path statistics, connected components, and inter-layer assortativity. This tool is an integral component of the comprehensive analytical suite developed by the CoMuNeLab~\citep{CoMuNeLabGithub}.

\begin{paragraph}{Tensor decomposition}
\texttt{muxvizpy} implements the Canonical Polyadic decomposition by reimplementing the Alternative Least Squares minimization algorithm with sparse operations~\citep{kolda2009tensor}. The algorithm involves the Matricized tensor times Khatri-Rao product (MTTKRP) and is validated against the python package \texttt{tensorly}~\citep{kossaifi2019tensorly} using pearson correlation or spearman correlation of the factors. The pseudocode and details are reported in Additional File ~1.
\end{paragraph}

\subsection{Benchmark setup}

Computational benchmarks compare \texttt{muxvizpy} and \texttt{muxViz} on multiplex networks generated using two standard random graph models: \emph{Erd\H{o}s--R\'enyi} (ER) and \emph{Barab\'asi--Albert} (BA). Detailed descriptions of network generation procedures and benchmark settings are provided in the Supplementary Material.

Benchmark experiments were performed for all metrics implemented in both \texttt{muxvizpy} and \texttt{muxViz}. Network sizes ranged from $10^5$ to $10^9$ edges, covering a broad spectrum of multilayer network dimensions. Each experiment was repeated three times on identical network instances to account for runtime variability.

For each run, wall-clock execution time and peak resident set size (RSS) memory were measured independently. Resource usage was collected through Slurm \texttt{jobacct-gather} utilities, while software-specific profiling tools were used to obtain fine-grained measurements: \texttt{bench} \citep{bench2025} for \texttt{muxViz}(R) and \texttt{tracemalloc} for \texttt{muxvizpy}(Python).
%% ================================================================
%% 3. RESULTS
%% ================================================================
\vspace{-8pt}
\section{Results}

\subsection{Computational benchmarks}
\label{sec:benchmarks}

\begin{figure*}[t]
%% PLACEHOLDER: replace with actual figure
\centerline{\includegraphics[height=0.4\textheight]{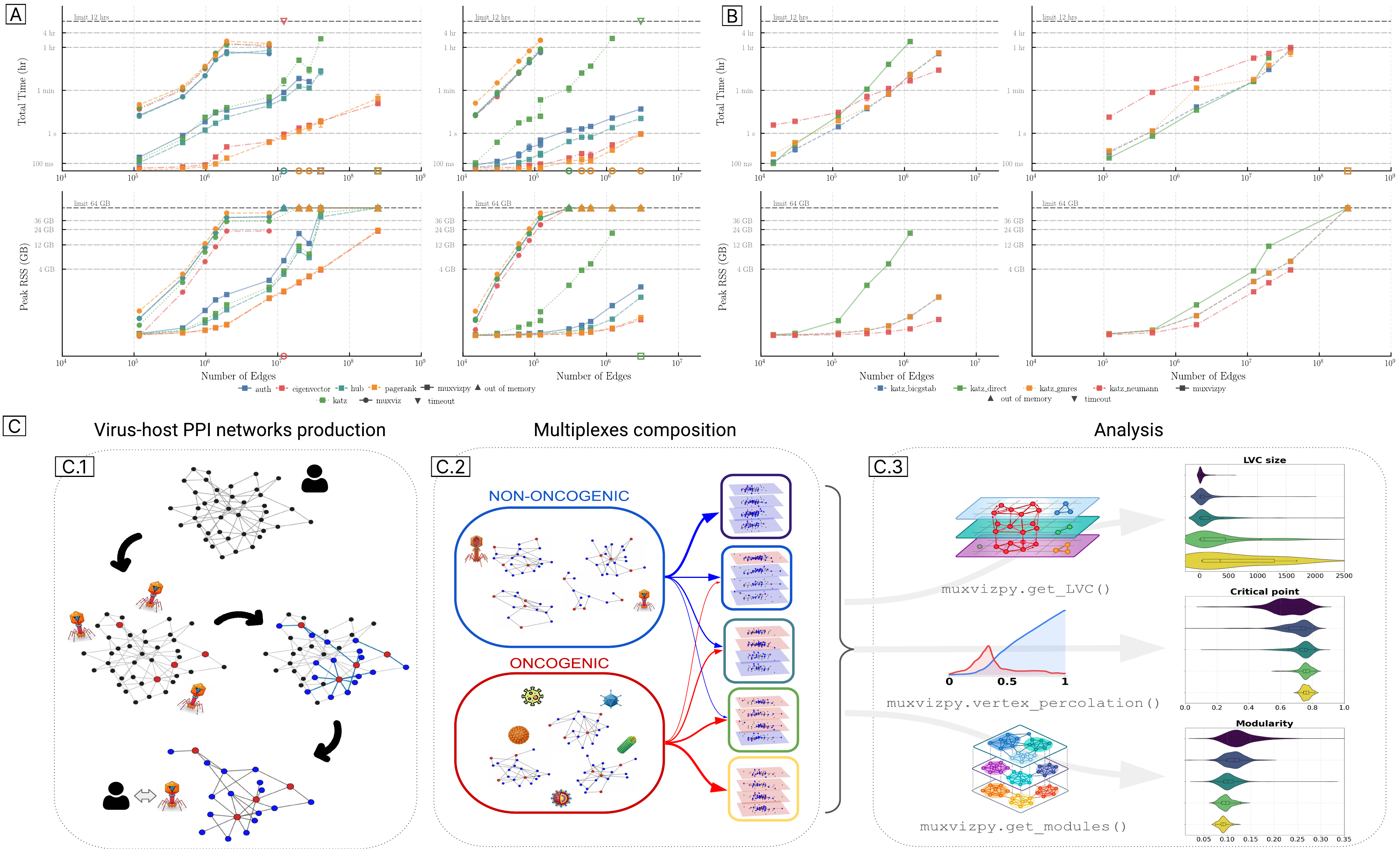}}
\caption{Computational cost comparison and usage application of muxvizpy. Panel A-B show the performances as a function of the number of edges $N$ for Erdős–Rényi (ER) on the first column and Barabási–Albert (BA) on the second column for multiplex networks with $L$ layers. Panel A displays versatility measures, while Panel B displays comparison between approximate and exact path to compute katz centrality. Colors refer to different metrics, while shape to the software. Each data point reflects the average of 3 replicates, with +- the standard deviation; markers for "timeout" and "out of memory" indicate instances where \texttt{muxViz} failed to complete the computation. Markers on the xaxis means that the run failed because of exceeding the costs of the other metric (so markers on time plots failed on oom and viceversa). Results were obtained on a Intel Xeon Platinum 8260 CPU @ 2.40/3.90GHz on an HPC cluster. (C) Workflow describing the biological use-case. C.1: Generation of protein-protein interaction (PPI) networks from curated datasets as proxies for human-virus interactions. C.2: Classification of viruses by oncogenic potential and generation of multiplex ensembles using varying proportions of oncogenic and non-oncogenic networks. C.3: Multilayer network analysis using \texttt{muxvizpy} functions (e.g., \texttt{get\_LVC()}, \texttt{vertex\_percolation()}, and \texttt{get\_modules()}) to extract metrics that provide a clear distinction between the two virus classes. }
\label{fig:benchmark}
\end{figure*}

% --- PLACEHOLDER: fill in with actual numbers from results/ directory ---
\texttt{muxvizpy} substantially extends the size of multiplex networks that can be analysed while preserving numerical agreement with \texttt{muxViz}. Although runtime does not increase strictly linearly with the number of edges, because it also depends on the convergence properties of the underlying spectral algorithms and on network topology, the observed scaling remains sufficiently favourable to extend the feasible network size by more than one order of magnitude. Accordingly, \texttt{muxvizpy} computed versatility measures on networks with up to approximately $10^{8}-10^{9}$ edges, whereas \texttt{muxViz} reached its time or memory limits near $10^{7}$ edges, as summarised in Fig \ref{fig:benchmark}. Eigenvector, PageRank, hub and authority centrality, completed the largest networks within seconds using sparse iterative eigensolvers. Katz centrality was more sensitive to network topology: the exact LU solver remained competitive up to approximately $10^{7}$ edges on ER networks but exceeded the memory limit beyond approximately $10^{6}$ edges on BA networks because of factorisation fill-in. In contrast, the ILU-preconditioned Krylov solvers continued to scale while retaining agreement with the exact solution. Across the shared versatility measures, \texttt{muxvizpy} achieved Pearson $r=1.0000$, with median mean absolute relative errors between $1.3\times10^{-4}$ and $8.0\times10^{-4}$. Computational results for the remaining metric categories are reported in Additional File~1.

\subsection{Use case: Virus-Human interaction multilayer PPI network}
As a representative biological application, we highlight a recent study by \cite{zambelli2025unraveling}, in which the core functionalities of \texttt{muxvizpy} were used to analyse multilayer virus--human protein--protein interaction networks. In this framework, each layer represented the interaction network associated with a different virus, while human proteins constituted the common substrate connecting layers. The resulting multiplexes contained approximately $10^4$ nodes and $10^5$ edges, placing them beyond the practical limits of several analyses available in existing multilayer software. The workflow is summarised in \ref{fig:benchmark}. First, virus-host PPI networks were reconstructed from curated interaction datasets and grouped according to the oncogenic or non-oncogenic nature of the corresponding viruses. These networks were then combined into multilayer ensembles, allowing the investigation of structural properties emerging from the joint organisation of multiple virus-specific interaction patterns. Finally, multilayer analyses including versatility measures, largest viable component (LVC) estimation, community detection, and vertex percolation were applied to characterise the resilience and mesoscale organisation of the system.

This application illustrates two practical advantages of \texttt{muxvizpy}. First, sparse data structures and iterative solvers enabled the computation of scalability-demanding metrics such as multilayer PageRank and other versatility measures on networks that could not be analysed efficiently with \texttt{muxViz}. Second, \texttt{muxvizpy} provided access to additional analyses, including percolation and tensor-based methods, that are not available in the original software. The resulting multilayer descriptors were sufficient to distinguish oncogenic from non-oncogenic viruses and helped identify candidate host genes associated with oncogenicity signatures, demonstrating the utility of large-scale multilayer network analysis for biological discovery.

\label{sec:usecase}

% --- PLACEHOLDER: describe an actual use case once available ---
%\tbd{}

%% ================================================================
%% 4. DISCUSSION AND CONCLUSIONS
%% ================================================================
\vspace{-12pt}
\section{Discussion}
We presented \textit{muxvizpy}, a Python library for multilayer network analysis that combines the analytical framework of\texttt{muxViz}with sparse tensor representations.
By representing the multilayer network as a four-dimensional sparse tensor and unfolding it on demand into a CSR supra-adjacency matrix, \texttt{muxvizpy} preserves the dimensional structure of the network and allows memory-efficient computation across seven feature categories spanning versatility, mesoscale, topology, information theory, tensor decomposition, global descriptors, and percolation.

Benchmark experiments on Erd\H{o}s--R\'enyi and Barab\'asi--Albert multiplex networks indicate numerical agreement with \texttt{muxViz} on shared metrics and show consistent reductions in wall-clock execution time and peak memory consumption across a range of network sizes. These results address limitations associated with dense network representations that have been previously reported for multilayer network software~\citep{Kjellstrom2023comparative, Buth2025}.

The virus--human PPI case study~\citep{zambelli2025unraveling} further demonstrates the applicability of \texttt{muxvizpy} to biologically relevant multilayer networks comprising approximately $10^4$ nodes and $10^5$ edges. In this setting, analyses based on PageRank versatility, percolation, and block-structure detection could be performed on networks for which equivalent analyses were either unavailable or computationally difficult to execute with \texttt{muxViz} because of memory requirements.

Two limitations should be noted. First, exact spectral and decomposition routines remain bounded by the cost of sparse eigensolvers; the approximate iterative paths exposed by \texttt{muxvizpy} mitigate but do not eliminate this constraint. Second, validation against \texttt{muxViz} is necessarily restricted to metrics implemented in both packages and to network sizes that \texttt{muxViz} can handle.

Future work will extend the decomposition module beyond CP to Tucker and block-term factorisations, add support to PyTorch Geometric, and develop bioinformatics-specific wrappers for AnnData and MuData so that multi-omics pipelines can invoke multilayer analyses without leaving the Scanpy ecosystem. By integrating natively with the Python scientific stack \texttt{muxvizpy} lowers the barrier to rigorous multilayer analysis of biological systems in which multiple interaction modalities must be modelled jointly.

%% ================================================================
%% FUNDING
%% ================================================================
\section*{Funding}

MB is funded by the European Union - Next Generation EU, Mission 4, Component 2, CUP B93D21010860004; F.Z. is supported by a Decreto Ministeriale n. 118 del 02/03/2023, M4C1 I. 4.1 from Piano Nazionale di Ripresa e Resilienza (PNRR). 

%% 
%% ACKNOWLEDGEMNTS
%%
\section*{Acknowledgements}
This work was supported in part by the high-performance computing infrastructure developed under the project “CONVECS”, funded by the PR Veneto FESR 2021-2027 program, Priority 1 – Specific Objective 1.1 – Action 1.1.2.

%% ================================================================
%% CONFLICT OF INTEREST
%% ================================================================
\vspace{-16pt}
\section*{Conflict of interest}

None declared.

%% ================================================================
%% REFERENCES
%% ================================================================
\vspace{-12pt}
\bibliographystyle{oup-abbrvnat}
\bibliography{reference}

%% file: supp.tex
\maketitle
\tableofcontents
\bigskip

%% ==================================================================
\section{Notation and data structures}
\label{sec:notation}

We consider a multilayer network with $N$ nodes and $L$ layers. Following the tensorial formalism of \citet{DeDomenico2013mathematical}, the fundamental object is the \emph{adjacency tensor} 
\begin{equation}
  \ten{M} \in \R^{N \times N \times L \times L}
\end{equation}
whose entry $\ten{M}_{ij\alpha\beta}$ is the weight of the directed edge from node $i$ in layer $\alpha$ to node $j$ in layer $\beta$. Intra-layer edges have $\alpha=\beta$ and inter-layer couplings have $\alpha\neq\beta$.
\texttt{muxvizpy} stores $\ten{M}$ as a PyTorch COO sparse tensor, preserving the node and layer dimensions explicitly.

For most structural metrics the tensor is flattened by a \emph{mode-1 unfolding} into the \emph{supra-adjacency matrix}
\begin{equation}
  \supra \in \R^{NL \times NL},
  \qquad
  \supra_{(\alpha-1)N + i,\; (\beta-1)N + j} = \ten{M}_{ij\alpha\beta},
  \label{eq:supra}
\end{equation}
stored as a SciPy CSR sparse matrix. $\supra$ is a block matrix whose $L$ diagonal blocks $\supra^{(\alpha)}\in\R^{N\times N}$ hold the intra-layer adjacencies and whose off-diagonal blocks hold the inter-layer couplings. 
For a node-aligned \emph{multiplex} network with categorical couplings, the off-diagonal blocks reduce to $\omega\,\mat{I}_N$ for a scalar coupling weight $\omega\ge 0$, which \texttt{muxvizpy} stores implicitly.

We write $\mat{I}$ for the identity, $\mathbf{1}$ for the all-ones vector, $\rho(\cdot)$ for the spectral radius, and $\lambda_{\max}(\supra)$ for the leading eigenvalue of $\supra$. A per-node \emph{versatility} score is obtained from a supra-vector $\mat{x}\in\R^{NL}$ by aggregating its layer replicas \citet{DeDomenico2015versatility} and max-normalising the result so that the largest score equals $1$:
\begin{equation}
  \mathrm{v}_i \;=\; \frac{s_i}{\max_{1\le j\le N} s_j},
  \qquad
  s_i \;=\; \sum_{\alpha=1}^{L} x_{(\alpha-1)N + i},
  \qquad i = 1,\dots,N.
  \label{eq:aggregate}
\end{equation}

%% ==================================================================
\section{Formal definition of the implemented metrics}
\label{sec:definitions}

We group the metrics by the seven categories of Table~1 in the main text and give each a self-contained definition. Unless we say otherwise, a definition acts on the supra-adjacency matrix $\supra$ of Eq.~\eqref{eq:supra}, and a per-node score is recovered from a node's layer replicas through Eq.~\eqref{eq:aggregate}.

\subsection{Versatility}
\label{sec:def-versatility}

\begin{definition}[Eigenvector versatility]
Eigenvector centrality calls a node important when its neighbours are important. Carried over to the supra-adjacency matrix, the score is its leading (Perron) eigenvector $\pi\in\R^{NL}$, solving $\supra\,\pi = \lambda_{\max}(\supra)\,\pi$ with $\pi\ge0$. The Perron--Frobenius theorem guarantees that such an eigenvector exists and can be chosen non-negative whenever $\supra$ is non-negative and irreducible.
\end{definition}

\begin{definition}[Katz versatility]
Katz versatility \citep{Katz1953,DeDomenico2015versatility} rewards a node for every walk that reaches it while discounting longer walks. It is also described as the generalisation of the eigenvector centrality for directed networks. In \texttt{muxvizpy} we use the form that includes the length-zero baseline term,
\begin{equation}
  \mat{x} \;=\; \sum_{k=0}^{\infty}\alpha^k\big(\supra^{\!\top}\big)^{k}\mathbf{1}
        \;=\; (\mat{I} - \alpha\supra^{\!\top})^{-1}\mathbf{1},
  \qquad \alpha \in (0,1/\lambda_{\max}(\supra)).
  \label{eq:katz}
\end{equation}
The series converges precisely when $\alpha\,\rho(\supra) < 1$, so \texttt{muxvizpy} pushes $\alpha$ as close to that limit as is safe, $\alpha = (1-\varepsilon)/\lambda_{\max}(\supra)$ for a small $\varepsilon>0$.
If a baseline-free Katz score is required, one subtracts $\mathbf{1}$ from Eq.~\eqref{eq:katz}; this constant shift does not affect node rankings in node-aligned multiplexes.
\end{definition}

\begin{definition}[PageRank versatility]
Picture a random surfer who at each step either follows an out-edge or, with probability $1-\alpha$, teleports to a node drawn from a personalisation vector $\mat{v}$. Writing $\mat{T} = \mat{D}^{-1}\supra$ for the row-stochastic transition matrix on non-dangling node-layers ($\mat{D}$ the diagonal out-strength matrix), with damping $\alpha\in(0,1)$ (default $0.85$) and $\mat{v}$ uniform by default, the surfer
moves under the Pagerank transition operator $\mat{G} = \alpha\mat{T} + (1-\alpha)\mathbf{1}\mat{v}^{\!\top}$, also called Google operator. Node-layers with zero out-strength are treated as dangling: their mass is redistributed according to $\mat{v}$, and the dangling indicator is denoted by $\mat{d}\in\{0,1\}^{NL}$ in Section~\ref{sec:pagerank-algo}. PageRank versatility is the share of time it spends at each node-layer pair in the long run, i.e.\ the stationary distribution $\boldsymbol{\pi}$ with $\mat{G}^{\!\top}\boldsymbol{\pi} = \boldsymbol{\pi}$, $\boldsymbol{\pi}\ge0$,
$\mathbf{1}^{\!\top}\boldsymbol{\pi}=1$.
\end{definition}

\begin{definition}[Random-walk (classical) versatility]
The same stationary idea without teleportation: $\boldsymbol{\pi}$ is the dominant left eigenvector of $\mat{T}$ at eigenvalue $1$, $\mat{T}^{\!\top}\boldsymbol{\pi} = \boldsymbol{\pi}$. Dropping the damping term also drops the guarantee of uniqueness --- on a multiplex that is not strongly connected the unit eigenvalue can be degenerate --- so we report this quantity for cost comparison only.
\end{definition}

\begin{definition}[Hub and authority versatility]
Following the HITS scheme, a good hub points to good authorities and a good authority is pointed to by good hubs. The two scores are the leading eigenvectors of $\supra\supra^{\!\top}$ and $\supra^{\!\top}\supra$ respectively, the leading left and right singular vectors of $\supra$ at its largest singular value $\sigma_1$. Each score is aggregated to physical nodes by Eq.~\eqref{eq:aggregate}.
\end{definition}

\begin{definition}[Multi-degree and multi-strength]
For a directed multiplex, degree is computed on the binary support of each intra-layer adjacency block, whereas strength keeps the edge weights. Unless stated otherwise, categorical inter-layer couplings are excluded from these per-layer degree and strength summaries. Thus, for layer $\alpha$,
\begin{equation}
  k^{\mathrm{in},(\alpha)}_i = \sum_j \mathbbm{1}\{\supra^{(\alpha)}_{ji} \neq 0\},
  \qquad
  k^{\mathrm{out},(\alpha)}_i = \sum_j \mathbbm{1}\{\supra^{(\alpha)}_{ij} \neq 0\},
\end{equation}
and the corresponding strengths replace the indicators by the weights $\supra^{(\alpha)}_{ji}$ and $\supra^{(\alpha)}_{ij}$. Aggregated in-, out- and total quantities are obtained by summing these layer-wise values over $\alpha$, with $k_i=k_i^{\mathrm{in}}+k_i^{\mathrm{out}}$.
\end{definition}

\begin{definition}[Multilayer $k$-core]
We compute the $k$-core on the unweighted support of the supra-adjacency graph: every non-zero intra- or inter-layer edge contributes one adjacency relation, irrespective of its weight. The algorithm repeatedly deletes node-layers of degree below $k$; what remains once the peeling stabilises is the $k$-core, the maximal sub-network in which every surviving node-layer keeps degree at least $k$. A node's
$k$-core centrality (coreness) is the largest $k$ for which one of its layer replicas survives the peeling.
\end{definition}

\subsection{Mesoscale}
\label{sec:def-mesoscale}

\begin{definition}[Multislice modularity]
Given a partition $\{g_{i\alpha}\}$ of the node-layers, the multislice modularity of \citet{Mucha2010community} scores how much more weight falls inside communities than a degree-preserving null model would put there,
\begin{equation}
  Q = \frac{1}{2\mu}\sum_{ij\alpha\beta}
      \Big[\big(\supra^{(\alpha)}_{ij} - \gamma_\alpha
      \tfrac{k_{i\alpha}k_{j\alpha}}{2m_\alpha}\big)\delta_{\alpha\beta}
      + \delta_{ij}\,C_{j\alpha\beta}\Big]\,
      \delta\!\big(g_{i\alpha},g_{j\beta}\big),
\end{equation}
where $m_\alpha$ is the edge count of layer $\alpha$, $k_{i\alpha}$ the degree of $i$ in $\alpha$, $\gamma_\alpha$ the resolution, $C_{j\alpha\beta}$ the coupling, and $2\mu = \sum_{j\alpha}(k_{j\alpha}+\sum_\beta C_{j\alpha\beta})$. Alongside modularity maximisation, \texttt{muxvizpy} exposes stochastic block model inference through \code{graph-tool} \citep{peixoto2023descriptive}.
\end{definition}

\begin{definition}[Inter-layer degree assortativity]
For a pair of layers $\alpha,\beta$, let $k_i^{(\alpha)}=k_i^{\mathrm{in},(\alpha)}+k_i^{\mathrm{out},(\alpha)}$ be the total intra-layer degree of node $i$ in layer $\alpha$ (or the analogous strength if weighted assortativity is requested). The inter-layer degree assortativity is the Pearson correlation of the two node-aligned sequences, $r_{\alpha\beta}=\operatorname{cov}(k^{(\alpha)},k^{(\beta)})/
(\sigma_{k^{(\alpha)}}\sigma_{k^{(\beta)}})$. A value near $1$ means a node that is a hub in $\alpha$ tends to be a hub in $\beta$ too, whereas a negative value signals the opposite, anti-correlated arrangement.
\end{definition}

\begin{definition}[Local clustering coefficient]
\citet{Cozzo2015structure} lift triadic closure to the multiplex by letting the three edges of a triangle live in different layers. Let $q_i$ be the number of connected multilayer triples centred on node $i$, counted over admissible layer assignments of the incident edges, and let $\tau_i$ be the subset of those triples that close into multilayer triangles. The local clustering coefficient is
\begin{equation}
  C_i =
  \begin{cases}
  \tau_i/q_i, & q_i>0,\\
  0, & q_i=0.
  \end{cases}
\end{equation}
\end{definition}

\subsection{Topology}
\label{sec:def-topology}

\begin{definition}[Largest connected component, LCC] The size, as a count or a fraction of node-layers, of the largest weakly connected component of the supra-adjacency graph.
\end{definition}

\begin{definition}[Largest intersection / viable component, LIC / LVC] The intersection graph keeps an edge $(i,j)$ only when it appears in \emph{every} layer, and the LIC is its largest connected component. The largest \emph{viable} component (LVC) is the maximal set of nodes that remain mutually reachable through within-layer paths in every layer after the iterative pruning of \citet{Baxter2012avalanche}.
\end{definition}

\begin{definition}[Path statistics and shortest-path similarity]
We summarise the distribution of multilayer shortest-path lengths $d(i\alpha,j\beta)$ measured on $\supra$ through the mean path length, efficiency and diameter. For node-level summaries we collapse layer replicas by the shortest available replica-to-replica path,
\begin{equation}
  d(i,j) = \min_{\alpha,\beta} d(i\alpha,j\beta),
\end{equation}
with unreachable pairs omitted from path-length averages and assigned zero efficiency. The shortest-path similarity matrix then uses the bounded monotone transform $S_{ij}=1/(1+d(i,j))$, with $S_{ij}=0$ when $i$ and $j$ are disconnected.
\end{definition}

\subsection{Information theory}
\label{sec:def-information}

\begin{definition}[Von Neumann entropy]
The construction borrows the density-matrix formalism of quantum mechanics.
From the combinatorial Laplacian $\mat{L}=\mat{D}-\supra$ one forms $\boldsymbol{\rho} = \mat{L}/\operatorname{tr}(\mat{L})$, which is positive-semidefinite and has unit trace, so it carries a von Neumann entropy 
\begin{equation}
  S(\boldsymbol{\rho}) = -\operatorname{tr}(\boldsymbol{\rho}\log\boldsymbol{\rho})
                       = -\sum_{k} \lambda_k \log \lambda_k,
  \label{eq:vn}
\end{equation}
taken over the eigenvalues $\{\lambda_k\}$ of $\boldsymbol{\rho}$ \citep{DeDomenico2016entropy}. This is the $Q\!\to\!1$ limit of the Rényi spectral entropy, and \texttt{muxvizpy} reproduces the \texttt{muxViz} value at $Q=1$ exactly.
\end{definition}

\begin{definition}[Jensen--Shannon divergence]
Given two density matrices $\boldsymbol{\rho},\boldsymbol{\sigma}$, the Jensen-Shannon divergence would be
\begin{equation}
  \mathrm{JSD}(\boldsymbol{\rho},\boldsymbol{\sigma}) =
  S\!\big(\tfrac{\boldsymbol{\rho}+\boldsymbol{\sigma}}{2}\big)
  - \tfrac{1}{2}\big[S(\boldsymbol{\rho}) + S(\boldsymbol{\sigma})\big]
\end{equation}
The Jensen-Shannon divergence is a symmetric, bounded measure of how differently the two are organised. Applied to every layer pair it fills an $L\times L$ matrix, at a cost of $O(L^2)$ dense eigendecompositions of $N\times N$ density matrices.
\end{definition}

\subsection{Decomposition}
\label{sec:def-decomposition}

\begin{definition}[Canonical Polyadic (CP/PARAFAC) decomposition]
The rank-$R$ CP decomposition summarises the adjacency tensor as a sum of $R$ rank-one terms, each with a single coherent pattern across nodes and layers:
\begin{equation}
  \ten{M} \approx \sum_{r=1}^{R} \lambda_r\,
    \mat{a}_r \circ \mat{b}_r \circ \mat{c}_r \circ \mat{d}_r,
  \label{eq:cp}
\end{equation}
where $\circ$ is the outer product, $\lambda_r$ is the component weight, and $\mat{a}_r,\mat{b}_r\in\R^N$, $\mat{c}_r,\mat{d}_r\in\R^L$ describe the source-node, target-node, source-layer and target-layer factors, respectively \citep{kolda2009tensor}. The sparse ALS algorithm used to fit these factors is defined in Section~\ref{sec:cp-algo}.
\end{definition}

\subsection{Global descriptors}
\label{sec:def-global}

\begin{definition}[Average global clustering coefficient]
A single scalar summarising triadic closure across the whole system: the mean of the local clustering coefficient (Section~\ref{sec:def-mesoscale}) over all node-layers.
\end{definition}

\begin{definition}[Global edge overlap and overlap matrices]
Two layers overlap to the extent that they share edges, which we quantify by the Jaccard index of their edge sets,
$o_{\alpha\beta} = |E_\alpha \cap E_\beta| / |E_\alpha \cup E_\beta|$ ($E_\alpha$ the edge set of layer $\alpha$). Collected over all pairs these values form the $L\times L$ edge overlap matrix, and averaging its off-diagonal entries gives the average global overlap. The node overlap matrix is built the same way from node participation across layer pairs.
\end{definition}

\subsection{Percolation}
\label{sec:def-percolation}

\begin{definition}[Vertex percolation]
Site percolation keeps each node independently with probability $p$ and asks how much of the multiplex still holds together. The order parameter is the relative size $P_\infty(p)$ of the largest viable component of the survivors; sweeping $p$ downward traces a robustness curve whose sharp drop locates the percolation threshold $p_c$. \texttt{muxvizpy} estimates $P_\infty(p)$ by repeatedly extracting the LVC over ensembles of random node removals.
\end{definition}

%% ==================================================================
\section{Exact and approximate algorithms}
\label{sec:algorithms}

Underneath their different definitions, several versatility measures come down to one of two linear-algebra problems: solving a linear system of the form $(\mat{I}-\alpha\supra^{\!\top})\mat{x}=\mathbf{1}$, or finding the dominant eigenvalue and related eigenvector. \texttt{muxvizpy} offers an \emph{exact} route, a sparse direct factorisation or a Krylov eigensolver with different possible preconditioning of the eigendecomposition. The bet is that past a certain size the exact route runs out of time or memory, while a well-chosen iterative scheme holds essentially the same accuracy
for a fraction of the cost\citep{NEURIPS2024_e88870ec,Saad2003iterative}, but without guarantees of exact solution, only guarantees on asymptotic convergence to the solution.

\subsection{Katz centrality}
\label{sec:katz-algo}

From Eq.~\eqref{eq:katz}, Katz centrality solves
\begin{equation}
  (\mat{I}-\alpha\supra^{\!\top})\mat{x} = \mathbf{1},
  \qquad \alpha = \frac{1-\varepsilon}{\lambda_{\max}(\supra)} .
  \label{eq:katz-system}
\end{equation}

\paragraph{Exact path (direct LU).}
\code{spsolve} forms a sparse Lower-Upper (LU) factorisation of $\mat{I}-\alpha\supra^{\!\top}$ with back-substitutes (Gauss factorisation), returning the exact solution. What makes it expensive is not the number of non-zeros in $\supra$ but the fill-in that the factorisation introduces in $\mat{L}$ and $\mat{U}$.

\paragraph{Approximate paths.}
\emph{(i) Neumann series.} Since $\alpha\,\rho(\supra)=1-\varepsilon<1$, the matrix $\mat{B}=\mat{I}-\alpha\supra^\top$ is nonsingular and its inverse admits the convergent Neumann expansion $\sum_{k\ge0}(\alpha\supra^\top)^k$. This gives a matrix-free fixed-point scheme, $\mat{x}\leftarrow\mathbf{1}+\alpha\supra^\top\mat{x}$, requiring only sparse matrix--vector products and no factorisation. Its convergence, however, is geometric with rate controlled by $\alpha\,\rho(\supra)=1-\varepsilon$; therefore, as $\alpha$ approaches the spectral limit $1/\rho(\supra)$, the iteration remains valid but becomes progressively slower. \emph{(ii) Krylov methods with Incomplete LU} (ILU) preconditioning, which is a technique in numerical linear algebra used to accelerate the convergence of iterative solvers for large, sparse linear systems.  It works by approximating the coefficient matrix $A$ with a product of lower and upper triangular matrices, $M \approx LU$, where fill-in elements are dropped to maintain sparsity.ILU preconditioning. In the same near-critical regime, $\mat{B}=\mat{I}-\alpha\supra^\top$ becomes increasingly ill-conditioned, since eigenvalues of $\alpha\supra^\top$ close to one are mapped close to zero. Krylov solvers such as GMRES and BiCGSTAB \cite{saad1985generalized,van1992bi} address this system by building approximations from residual-generated subspaces, rather than by explicitly accumulating the Neumann powers. Their convergence is improved by an ILU preconditinoer $\mat{M}\approx\mat{B}$, which replaces the original system by the preconditioned one $\mat{M}^{-1}\mat{B}\mat{x}=\mat{M}^{-1}\mathbf{1}$. When the ILU factors provide a sparse but effective approximation of $\mat{B}$, the preconditioned operator has a more favourable spectrum and the residual typically decreases in far fewer iterations. This adds a one-time setup cost and extra memory for the incomplete factors, controlled by the fill level or drop tolerance, but substantially reduces the number of sparse matrix--vector products in the ill-conditioned regime. For this reason, ILU-preconditioned Krylov solvers are the preferred scalable option when $\alpha$ is chosen close to the spectral limit.

\begin{algorithm}[ht]
\caption{Katz centrality --- exact and approximate paths}
\label{alg:katz}
\begin{algorithmic}[1]
\Require supra-adjacency $\supra$, attenuation gap $\varepsilon$, residual tolerance $\mathrm{tol}$, maxiter, mode
         $\in\{\textsc{direct},\textsc{neumann},\textsc{gmres},\textsc{bicgstab}\}$
\State $\lambda_{\max}\gets$ \Call{LeadingEigenvalue}{$\supra$} \Comment{ARPACK, $k{=}1$}
\State $\alpha \gets (1-\varepsilon)/\lambda_{\max}$;\quad
       $\mat{B}\gets \mat{I}-\alpha\,\supra^{\!\top}$
\If{mode $=$ \textsc{direct}} \Comment{exact}
  \State $\mat{x}\gets$ \Call{SparseLUSolve}{$\mat{B},\ \mathbf{1}$}
\ElsIf{mode $=$ \textsc{neumann}} \Comment{approximate, unconditional}
  \State $\mat{x}\gets\mathbf{1}$
  \Repeat
    \State $\mat{x}\gets \mathbf{1}+\alpha\,\supra^{\!\top}\mat{x}$
  \Until{$\lVert\mat{B}\mat{x}-\mathbf{1}\rVert < \mathrm{tol}$ \textbf{or} maxiter}
\Else \Comment{approximate, Krylov + ILU}
  \State $\mat{P}\gets$ \Call{ILU}{$\mat{B}$};\quad
         $\mat{x}\gets$ \Call{Krylov}{$\mat{B},\ \mathbf{1},\ \mat{P}$}
         \Comment{GMRES or BiCGSTAB}
\EndIf
\State \Return per-node aggregation of $\mat{x}$ \Comment{Eq.~\eqref{eq:aggregate}}
\end{algorithmic}
\end{algorithm}

\subsection{PageRank and random-walk centrality}
\label{sec:pagerank-algo}

Both scores are stationary distributions of a walk on the row-stochastic transition matrix $\mat{T}$. Given the properties of such transition matrix, the Perron--Frobenius \citep{Langville2006google} theorem holds and guarantees that a non-negative irreducible stochastic matrix has a unique such distribution $\boldsymbol{\pi}$, namely the dominant eigenvector of $\mat{T}^{\!\top}$ at $\lambda_0=1$; the two metrics differ only in how they reach it. Therefore, solving an asymptotic series by power iteration of the dynamic phenomenon of random walkers or decomposing in the spectral properties of the transition matrix yields the same stationary distribution.

\paragraph{Exact path (eigensolver).}
The classical random-walk centrality reads $\boldsymbol{\pi}$ straight off a sparse eigendecomposition (\code{eigs} with $k=1$): internally this builds an Krylov factorisation and returns the dominant eigenpair, without ever iterating the walk explicitly. Again, we can precondition the eigensolver since the transition matrix has similar properties to the ones described above.

\paragraph{Approximate path (power iteration).}
PageRank regularises the operator by teleportation, $\mat{G}=\alpha\mat{T}+(1-\alpha)\mathbf{1}\mat{v}^{\!\top}$, which makes every
entry strictly positive so that Perron--Frobenius applies even when the graph is reducible or carries dangling node-layers. The stationary vector is then the limit of the power iteration
\begin{equation}
  \mat{x}_{k+1} = \alpha\,\mat{T}^{\!\top}\mat{x}_k
                 + \big(\alpha\,\mat{d}^{\!\top}\mat{x}_k + 1 - \alpha\big)\,\mat{v},
  \label{eq:pagerank-iter}
\end{equation}
with $\mat{d}$ the dangling-node indicator and an $\ell_1$ renormalisation after each step. Convergence is geometric at ratio $\alpha$: teleportation widens the spectral gap $1-\lvert\lambda_1\rvert$, so even $\alpha=0.85$ already yields fast, robust convergence.

\begin{algorithm}[ht]
\caption{PageRank / random-walk centrality --- exact and approximate paths}
\label{alg:pagerank}
\begin{algorithmic}[1]
\Require supra-adjacency $\supra$, damping $\alpha$, personalisation $\mat{v}$, mode
\State $\mat{T}\gets \mat{D}^{-1}\supra$;\quad
       $\mat{d}\gets$ dangling-node mask
\If{mode $=$ \textsc{exact}} \Comment{eigensolver}
  \State $\boldsymbol{\pi}\gets$ \Call{Eigs}{$\mat{T}^{\!\top},\ k{=}1$}
         \Comment{dominant eigenpair, $\lambda_0{=}1$}
\Else \Comment{approximate, power iteration}
  \State $\mat{x}\gets \mat{v}$
  \Repeat
    \State $\mat{x}\gets \alpha\,\mat{T}^{\!\top}\mat{x}
            + (\alpha\,\mat{d}^{\!\top}\mat{x} + 1-\alpha)\,\mat{v}$;\quad
           $\mat{x}\gets \mat{x}/\lVert\mat{x}\rVert_1$
  \Until{$\lVert\mat{x}_{k+1}-\mat{x}_k\rVert_1 < \mathrm{tol}$ \textbf{or} maxiter}
  \State $\boldsymbol{\pi}\gets\mat{x}$
\EndIf
\State \Return per-node aggregation of $\boldsymbol{\pi}$
\end{algorithmic}
\end{algorithm}

\subsection{Sparse CP/PARAFAC decomposition}
\label{sec:cp-algo}

The rank-$R$ CP decomposition of Section~\ref{sec:def-decomposition} factorises the four-way adjacency tensor $\ten{M}\in\R^{N\times L\times N\times L}$ into factor matrices $\mat{A}\in\R^{N\times R}$, $\mat{B}\in\R^{L\times R}$, $\mat{C}\in\R^{N\times R}$, $\mat{D}\in\R^{L\times R}$ and weights $\boldsymbol{\lambda}\in\R^{R}$ such that $\ten{M}\approx\sum_{r=1}^{R}\lambda_r\,
    \mat{a}_r\circ\mat{b}_r\circ\mat{c}_r\circ\mat{d}_r$.
Fitting all four factors at once is non-convex, but it has a redeeming structure that \emph{alternating least squares} (ALS) exploits: with any three factors held fixed, the objective $\lVert\ten{M}-\hat{\ten{M}}\rVert_F^2$ is merely quadratic in the fourth \citep{kolda2009tensor}. \texttt{muxvizpy} therefore cycles through the four modes, and each sub-step collapses to an ordinary linear least-squares solve.

\paragraph{Mode-$n$ update.}
Writing the four factor matrices as $\mat{A}^{(1)}=\mat{A}$, $\mat{A}^{(2)}=\mat{B}$, $\mat{A}^{(3)}=\mat{C}$ and $\mat{A}^{(4)}=\mat{D}$, the
least-squares-optimal update of the mode-$n$ factor with the rest held fixed is
\begin{equation}
  \mat{A}^{(n)} \;\leftarrow\;
  \ten{M}_{(n)}\,\mat{K}_n\,\big(\mat{V}_n\big)^{\dagger},
  \qquad
  \mat{V}_n \;=\; \mathop{\scalebox{1.2}{$\ast$}}_{m\neq n}
                  \big(\mat{A}^{(m)\top}\mat{A}^{(m)}\big),
  \label{eq:cp-update}
\end{equation}
where $\ten{M}_{(n)}$ is the mode-$n$ unfolding, $\mat{K}_n$ the Khatri--Rao product, also called Kronecker product,  of the remaining factors in descending mode order, $\ast$ the Hadamard product, $(\cdot)^{\dagger}$ the Moore--Penrose pseudo-inverse of the $R\times R$ Gram matrix $\mat{V}_n$, and $\ten{M}_{(n)}\mat{K}_n$ is the matricised-tensor-times-Khatri--Rao product (MTTKRP). The dense Khatri--Rao factor $\mat{K}_n$ is never formed: because $\ten{M}$ is stored as a COO sparse tensor, the MTTKRP is evaluated directly from its non-zeros, so the per-iteration ost scales with $\mathrm{nnz}(\ten{M})\,R$ rather than with the tensor volume. After each solve we normalise the columns of $\mat{A}^{(n)}$ and push their norms into $\boldsymbol{\lambda}$, tracking convergence through the relative fit $1-\lVert\ten{M}-\hat{\ten{M}}\rVert_F/\lVert\ten{M}\rVert_F$.

\begin{algorithm}[ht]
\caption{Sparse CP/PARAFAC decomposition by alternating least squares}
\label{alg:cp-als}
\begin{algorithmic}[1]
\Require adjacency tensor $\ten{M}$ (COO sparse), rank $R$, tolerance $\mathrm{tol}$, maxiter
\State initialise factors $\{\mat{A}^{(1)},\mat{A}^{(2)},\mat{A}^{(3)},\mat{A}^{(4)}\}$
       \Comment{random or higher-order SVD (HOSVD), columns $\ell_2$-normalised}
\State $\mathrm{fit}\gets 0$
\Repeat
  \State $\mathrm{fit}_{\mathrm{old}}\gets\mathrm{fit}$
  \For{$n \in \{1,2,3,4\}$} \Comment{one mode at a time}
    \State $\mat{V}_n \gets \mathop{\scalebox{1.1}{$\ast$}}_{m\neq n}\big(\mat{A}^{(m)\top}\mat{A}^{(m)}\big)$
           \Comment{Hadamard of $R\times R$ Gram matrices}
    \State $\mat{G}_n \gets$ \Call{SparseMTTKRP}{$\ten{M},\ \{\mat{A}^{(m)}\}_{m\neq n}$}
           \Comment{$\ten{M}_{(n)}\mat{K}_n$ from non-zeros}
    \State $\mat{A}^{(n)} \gets \mat{G}_n\,\mat{V}_n^{\dagger}$
    \State $\boldsymbol{\lambda}\gets$ column norms of $\mat{A}^{(n)}$;\quad
           normalise columns of $\mat{A}^{(n)}$
  \EndFor
  \State $\mathrm{fit}\gets 1-\lVert\ten{M}-\hat{\ten{M}}\rVert_F/\lVert\ten{M}\rVert_F$
         \Comment{$\hat{\ten{M}}=\sum_r\lambda_r\,\mat{a}_r\!\circ\!\mat{b}_r\!\circ\!\mat{c}_r\!\circ\!\mat{d}_r$}
  \Until{$\lvert\mathrm{fit}-\mathrm{fit}_{\mathrm{old}}\rvert<\mathrm{tol}$ \textbf{or} maxiter}
\State \Return weights $\boldsymbol{\lambda}$ and factors $\{\mat{A}^{(1)},\mat{A}^{(2)},\mat{A}^{(3)},\mat{A}^{(4)}\}$
\end{algorithmic}
\end{algorithm}

%% ==================================================================
\section{Extended benchmark analysis}
\label{sec:benchmarks}

\subsection{Network generation and profiling protocol}
\label{sec:protocol}

Benchmarks compare \texttt{muxvizpy} and \texttt{muxViz} on synthetic multiplex networks whose $L$ layers are independent realisations of two random-graph models: \emph{Erd\H{o}s--R\'enyi} (ER), with connection probability $p$ controlling a homogeneous Poisson degree distribution, and
\emph{Barab\'asi--Albert} (BA), with attachment parameter $m$ giving a scale-free, heavy-tailed degree distribution. Networks are node-aligned
multiplexes with categorical inter-layer coupling. Edge lists are 0-indexed, directed CSV files (\code{node.from,\,layer.from,\,node.to,\,layer.to,\,weight}) shared byte-for-byte by both tools.

For each metric we sweep the number of nodes $N$ and the number of layers $L\in\{5,7,10\}$, spanning roughly $10^4$ to $2.5\times10^{8}$ edges. Each (software, metric, network, replicate) run is dispatched as an isolated SLURM job and repeated three times. Compute time (excluding network loading) and peak resident set size (RSS) are recorded independently: through \code{tracemalloc} and a dedicated profiler for \texttt{muxvizpy}, and through \code{bench} \citep{bench2025} and \code{/proc/self/status} for \texttt{muxViz}. Runs exceeding the job time or memory limits are retained as scalability ceilings. All results were obtained on Intel Xeon Platinum 8260 CPUs (2.40/3.90\,GHz).

In every cost figure, filled squares denote \texttt{muxvizpy} and open circles denote \texttt{muxViz}; one colour is used per metric. Horizontal dashed lines mark wall-clock thresholds (1\,min to 12\,h) and memory thresholds; the solid red line marks the per-job memory ceiling (\code{OOM}). Reference lines are shown only within each panel's data range, so the vertical resolution adapts to the cost regime of each metric category.

\subsection{Versatility centralities}
\label{sec:bench-versatility}
\begin{figure}[htbp]
\centering
\includegraphics[width=0.8\textwidth]{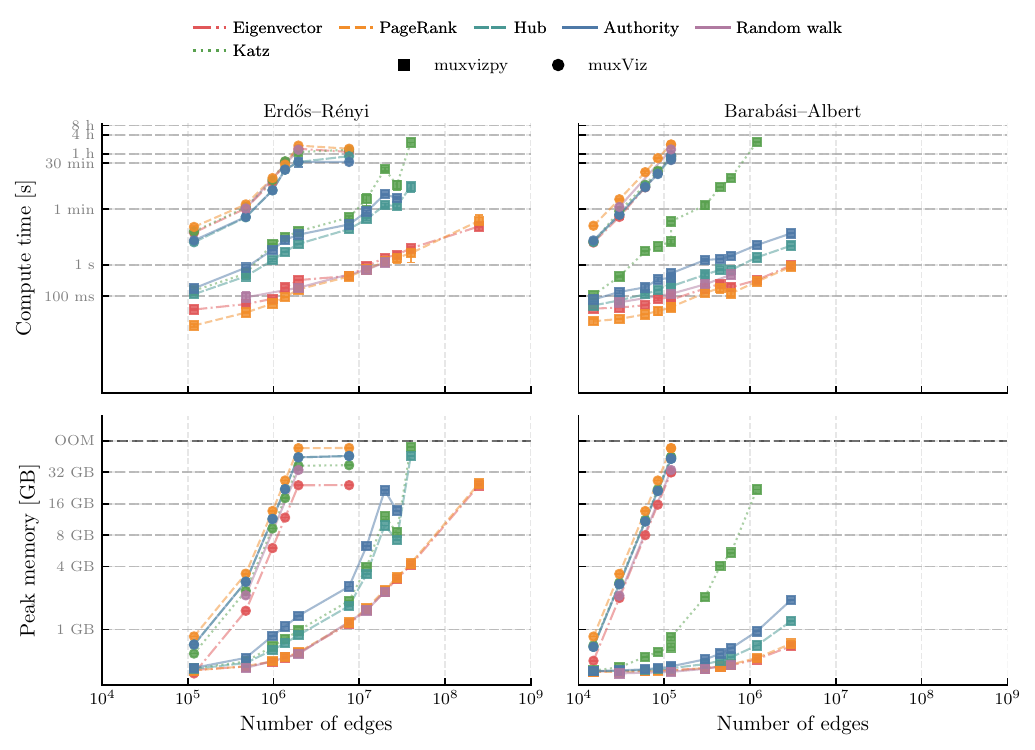}
\caption{Cost of the versatility centralities on ER (left) and BA (right). Top: compute time; bottom: peak memory, versus number of edges (log--log). Filled squares: \texttt{muxvizpy}; open circles: \texttt{muxViz}.}
\label{fig:bench-versatility}
\end{figure}

Figure~\ref{fig:bench-versatility} reports the eight versatility measures. The spectral centralities (eigenvector, PageRank, hub, authority) are solved with sparse iterative eigensolvers and complete ER multiplexes up to $\sim\!2.5\times10^8$ edges in seconds, whereas \texttt{muxViz} exceeds time or memory limits already around $10^7$ edges. Katz is the most expensive versatility metric because of the resolvent solve (Section~\ref{sec:katz-bench}); in-degree, out-degree and the random walk (shown where available) follow the cheap-kernel and spectral regimes respectively.

\subsection{Topology}
\label{sec:bench-topology}
\begin{figure}[htbp]
\centering
\includegraphics[width=0.8\textwidth]{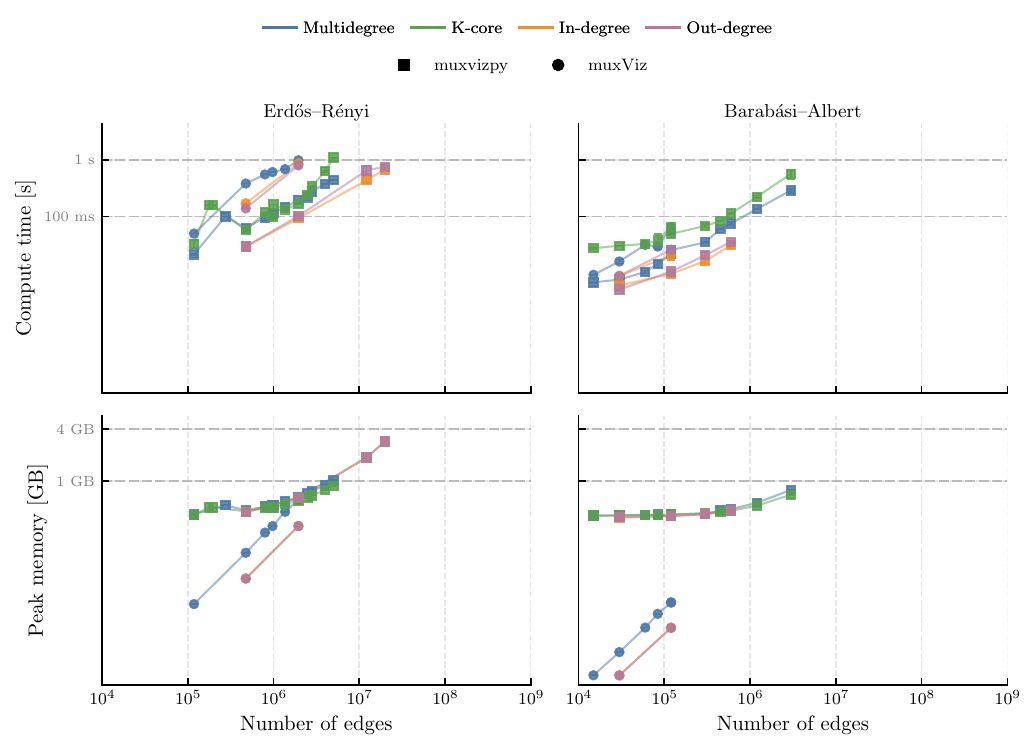}
\caption{Cost of the topology metrics on ER (left) and BA (right). Note the sub-second time axis: these metrics never approach the 1-minute
reference, so no wall-clock reference line falls within range.} 
\label{fig:bench-topology}
\end{figure}

Figure~\ref{fig:bench-topology} shows the combinatorial topology metrics (multidegree, $k$-core, in/out-degree). These are inexpensive in both tools (sub-second across the tested range); \texttt{muxvizpy} extends the feasible range by roughly one order of magnitude in edges. The apparent memory disadvantage of \texttt{muxvizpy} at the smallest sizes is the fixed $\sim\!0.4$\,GB Python/PyTorch import baseline, not worse scaling.

\subsection{Global descriptors}
\label{sec:bench-global}
\begin{figure}[htbp]
\centering
\includegraphics[width=0.8\textwidth]{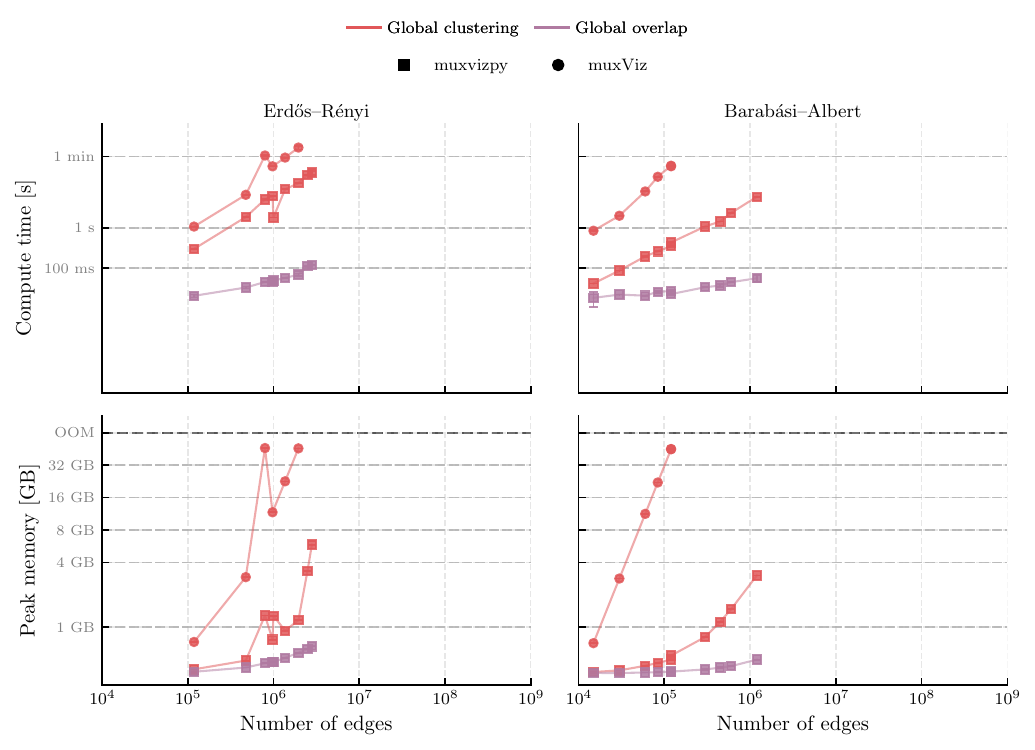}
\caption{Cost of the global descriptors on ER (left) and BA (right).}
\label{fig:bench-global}
\end{figure}

Figure~\ref{fig:bench-global} contrasts the two global descriptors. Global clustering, which requires a sparse $\supra^3$-type triangle count on the $(NL)\times(NL)$ supra-matrix, is the dominant cost and shows the widest gap: \texttt{muxViz} approaches its $\sim\!45$\,GB memory ceiling already at $\sim\!10^5$ (BA) to $\sim\!2\times10^6$ (ER) edges, while \texttt{muxvizpy} stays one-to-two orders of magnitude lower in memory. Global edge overlap is near-free.

\subsection{Mesoscale}
\label{sec:bench-mesoscale}
\begin{figure}[htbp]
\centering
\includegraphics[width=0.8\textwidth]{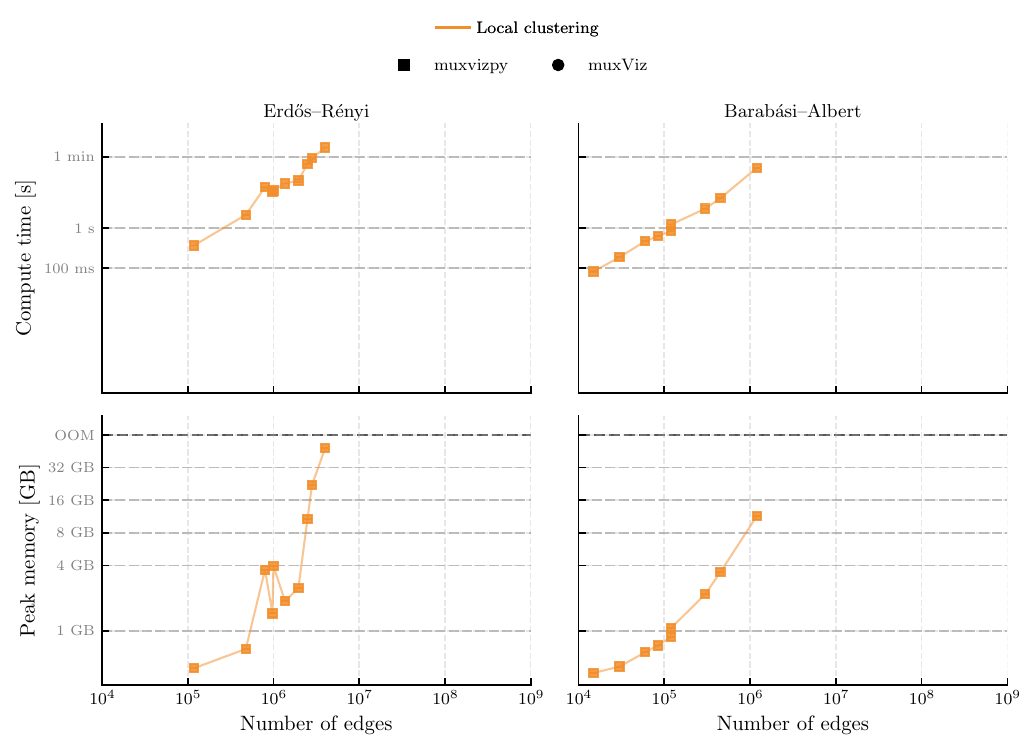}
\caption{Cost of the mesoscale local clustering coefficient on ER (left) and BA (right). \texttt{muxvizpy}-only (\texttt{muxViz} values were not retained for this metric).}
\label{fig:bench-mesoscale}
\end{figure}

Figure~\ref{fig:bench-mesoscale} reports the local clustering coefficient, the mesoscale metric exercised in the benchmark. Like global clustering it is triangle-bound; \texttt{muxvizpy} completes ER multiplexes up to $4\times10^6$ edges before approaching the memory ceiling. Community detection and inter-layer assortativity are defined in Section~\ref{sec:def-mesoscale} but are exercised in the biological use case of the main text rather than in this synthetic sweep, as they require a \code{graph-tool} graph and a per-layer dictionary output, respectively, that fall outside the uniform benchmark harness.

\subsection{Information theory}
\label{sec:bench-information}
\begin{figure}[htbp]
\centering
\includegraphics[width=0.8\textwidth]{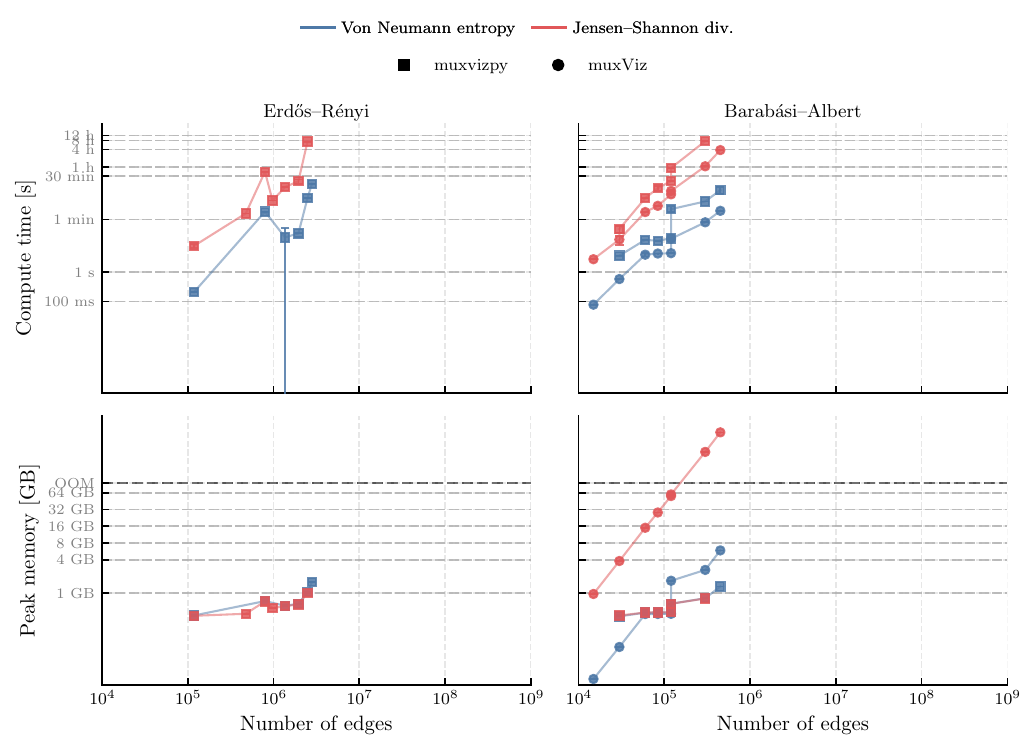}
\caption{Cost of the information-theoretic metrics on ER (left) and BA (right).}
\label{fig:bench-information}
\end{figure}

Figure~\ref{fig:bench-information} reports von Neumann entropy and layer-pairwise Jensen--Shannon divergence. Their cost is set by dense eigendecompositions of $N\times N$ density matrices ($O(L^2)$ of them for JSD), making this a small/medium-size category in both tools. The advantage of \texttt{muxvizpy} here is concentrated in memory: on BA the pairwise JSD stays near $0.5$\,GB versus $\sim\!15$\,GB for \texttt{muxViz} ($\sim\!30\times$ less), at comparable wall-clock time.

\subsection{Decomposition}
\label{sec:bench-decomposition}
\begin{figure}[htbp]
\centering
\includegraphics[width=0.8\textwidth]{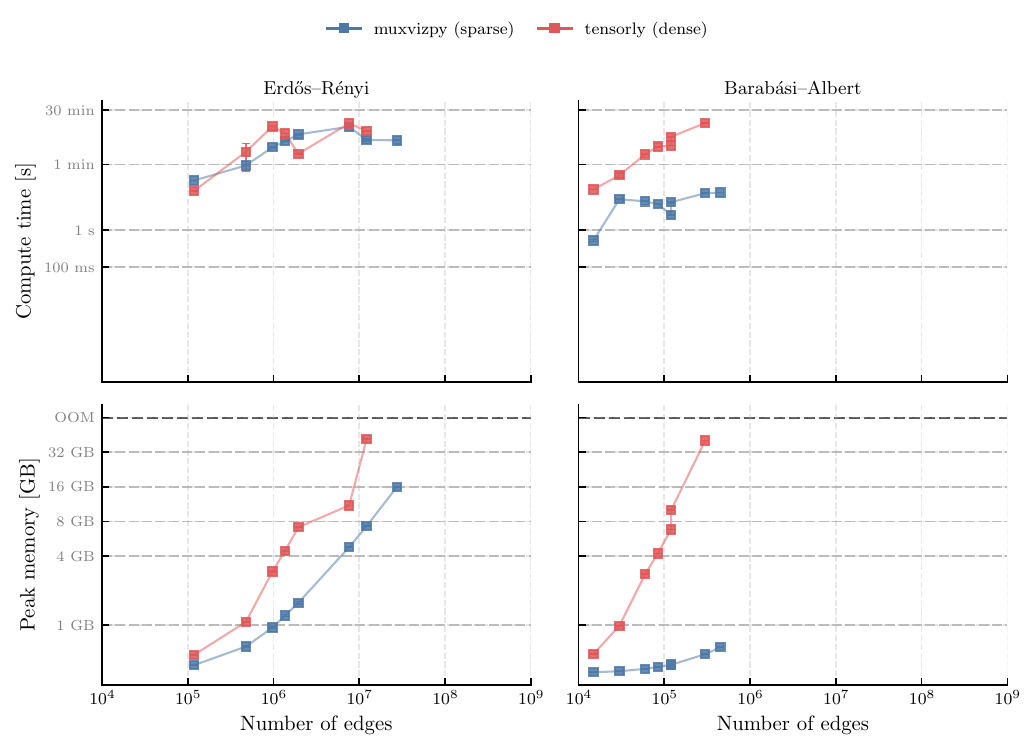}
\caption{Cost of the CP/PARAFAC decomposition (rank $R=10$) on ER (left) and BA (right): \texttt{muxvizpy}'s sparse ALS versus \code{tensorly}'s dense \code{parafac}. Top: compute time; bottom: peak memory, versus number of edges (log--log). \texttt{muxvizpy}'s cost tracks $\mathrm{nnz}(\ten{M})$, whereas \code{tensorly} densifies the $N^2L^2$ tensor and reaches the time/memory ceilings at moderate sizes. Points are medians over replicates.}
\label{fig:bench-decomposition}
\end{figure}

Figure~\ref{fig:bench-decomposition} compares the cost of the sparse CP/PARAFAC decomposition (Section~\ref{sec:cp-algo}) with the dense reference implementation in \code{tensorly} \citep{kossaifi2019tensorly}---the same reference used for the factor validation of Section~\ref{sec:acc-decomposition}---both at fixed rank $R=10$ and the same stopping criterion ($\mathrm{tol}=10^{-6}$, at most $100$ ALS sweeps). The two solve the identical optimisation problem but pay very different costs. \texttt{muxvizpy} evaluates the MTTKRP directly from the non-zeros, so each sweep costs $\mathrm{nnz}(\ten{M})\,R$; its peak memory grows smoothly with the edge count---from $\sim\!0.4$\,GB to $\sim\!16$\,GB across the ER range---and it completes even the densest $\sim\!2.7\times10^7$-edge multiplexes within the per-job budget. \code{tensorly}'s \code{parafac}, by contrast, must materialise the dense $N\times L\times N\times L$ tensor and its mode unfoldings, so its cost tracks the tensor \emph{volume} $N^2L^2$ regardless of sparsity: it is competitive only on the smallest networks and then runs into the time and memory limits at moderate sizes, well before \texttt{muxvizpy} shows any strain. Neither wall-clock curve is a pure function of size, since the number of sweeps to reach the tolerance is data-dependent. Each point is the median over replicates; this metric has no \texttt{muxViz} counterpart.

\subsection{Percolation}
\label{sec:bench-percolation}
\begin{figure}[ht]
\centering
\includegraphics[width=0.8\textwidth]{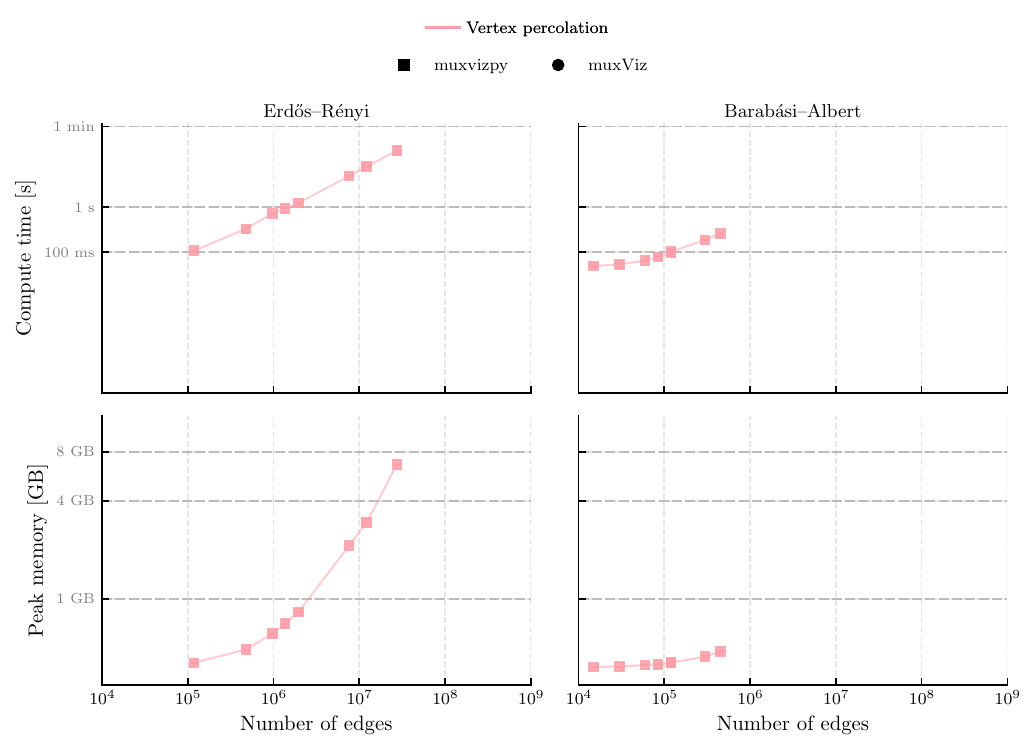}
\caption{Cost of vertex percolation on ER (left) and BA (right). Top: compute time; bottom: peak memory, versus number of edges (log--log).
\texttt{muxvizpy}-only (no \texttt{muxViz} counterpart).}
\label{fig:bench-percolation}
\end{figure}

Figure~\ref{fig:bench-percolation} reports the cost of vertex percolation (Section~\ref{sec:def-percolation}). Since the order parameter $P_\infty(p)$ is estimated by repeated LVC extraction over random node-removal ensembles, the cost is set by the length of the $p$-sweep times the price of a single viable-component computation, making it the most expensive topology-side metric per network.

\subsection{Exact versus approximate Katz solvers}
\label{sec:katz-bench}
\begin{figure}[ht]
\centering
\includegraphics[width=\textwidth]{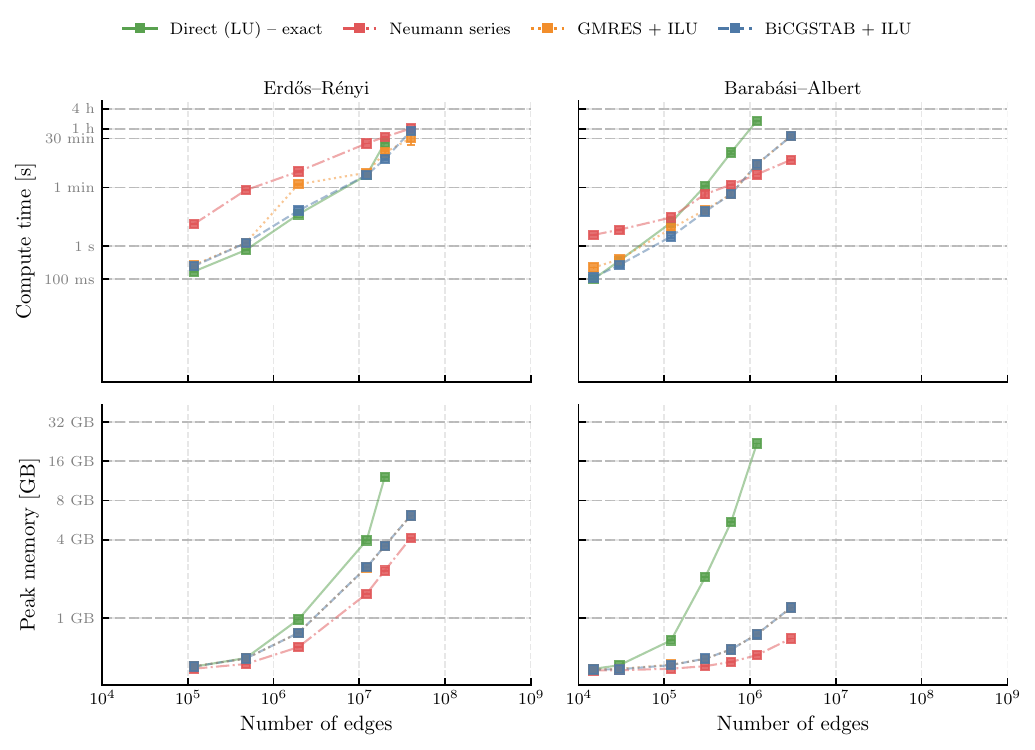}
\caption{Exact (direct LU) versus approximate (Neumann, GMRES$+$ILU, BiCGSTAB$+$ILU) Katz solvers. On ER the direct solver is competitive up to $\sim\!10^7$ edges before Krylov solvers overtake it; on BA the direct solver suffers catastrophic hub-induced LU fill-in and runs out of memory beyond $\sim\!10^6$ edges, whereas the preconditioned Krylov and Neumann paths continue to scale.}
\label{fig:katz-solvers}
\end{figure}

Figure~\ref{fig:katz-solvers} contrasts the four Katz solver paths of Algorithm~\ref{alg:katz}. The choice of solver, not the algorithm, governs scalability, and the exact direct path degrades sharply on scale-free topologies.

\clearpage
%% ==================================================================
\section{Numerical accuracy}
\label{sec:accuracy}

\begin{figure}[htbp]
\centering
\includegraphics[width=\textwidth]{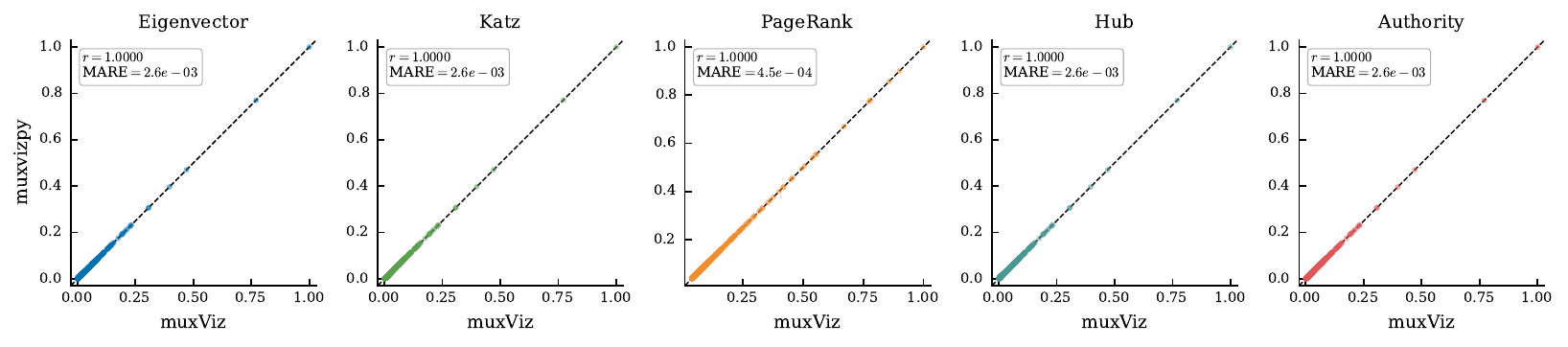}
\caption{Node-level agreement between \texttt{muxvizpy} ($y$) and \texttt{muxViz} ($x$) for the versatility centralities at the largest shared network. The dashed line is the identity; the box reports Pearson $r$ and MARE.}
\label{fig:acc-versatility}
\end{figure}

\begin{figure}[htbp]
\centering
\includegraphics[width=0.72\textwidth]{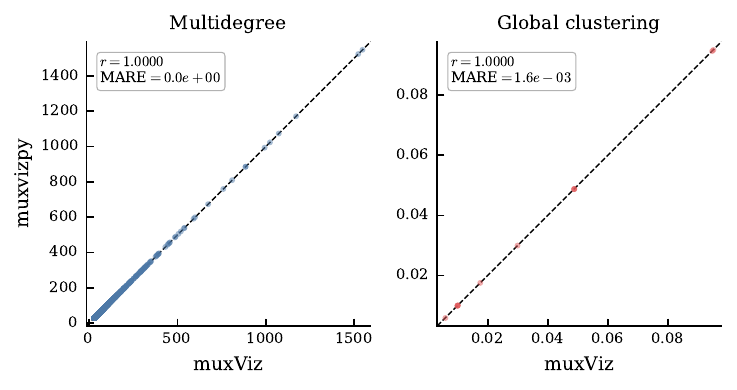}
\caption{Agreement between \texttt{muxvizpy} ($y$) and \texttt{muxViz} ($x$) for the remaining paired metrics. Left: per-node multidegree at the largest shared network, which matches exactly ($r=1.0000$, MARE $0$). Right: the global clustering coefficient, one scalar per network, across all $12$ benchmark networks ($r=1.0000$, MARE $\sim\!10^{-3}$). The dashed line is the identity.}
\label{fig:acc-topology-global}
\end{figure}

\texttt{muxViz} is treated as ground truth; accuracy is therefore evaluated only for metrics computed by both tools and for network sizes \texttt{muxViz} can handle. For each shared metric and each benchmark network we pair per-node values (or the scalar, for global metrics) and report the Pearson correlation $r$, together with the mean absolute relative error (MARE). Table~\ref{tab:accuracy} summarises the results: all versatility centralities agree with \texttt{muxViz} to machine precision ($r=1.0000$; MARE $\sim\!10^{-4}$), multidegree is exact, the global clustering scalar agrees to $\sim\!10^{-3}$, and the information-theoretic metrics agree to $\sim\!10^{-4}$ or better (BA only; Section~\ref{sec:acc-information}). Figures~\ref{fig:acc-versatility} and~\ref{fig:acc-topology-global} show the corresponding node- and network-level scatter at the largest shared networks.

\begin{table}[ht]
\centering
\caption{Numerical agreement with \texttt{muxViz} across all benchmark networks. $r$ is Pearson correlation of paired per-node values (n/a for the single-scalar global clustering and von Neumann entropy); MARE is the mean absolute relative error. Ranges are over the benchmark network sizes ($n$ networks). $^{\dagger}$The information-theoretic metrics are compared on the BA multiplexes only, since \texttt{muxViz} fails on the ER instances.}
\label{tab:accuracy}
\small
\begin{tabular}{@{}l l c c c@{}}
\toprule
\textbf{Category} & \textbf{Metric} & \textbf{$n$} & \textbf{$r$ (min)} & \textbf{MARE (median / max)} \\
\midrule
Versatility  & Eigenvector       & 12 & $1.0000$ & $3.9\times10^{-4}$ / $5.8\times10^{-3}$ \\
Versatility  & Katz              & 12 & $1.0000$ & $8.0\times10^{-4}$ / $5.7\times10^{-3}$ \\
Versatility  & PageRank          & 12 & $1.0000$ & $1.3\times10^{-4}$ / $6.9\times10^{-4}$ \\
Versatility  & Hub               & 12 & $1.0000$ & $3.2\times10^{-4}$ / $5.8\times10^{-3}$ \\
Versatility  & Authority         & 12 & $1.0000$ & $2.6\times10^{-4}$ / $5.8\times10^{-3}$ \\
Topology     & Multidegree       & 12 & $1.0000$ & $0$ / $0$ (exact) \\
Global       & Global clustering & 12 & n/a      & $1.0\times10^{-3}$ / $7.4\times10^{-3}$ \\
Info.\ theory & Von Neumann entropy$^{\dagger}$   & 4 & n/a      & $2.7\times10^{-6}$ / $6.7\times10^{-6}$ \\
Info.\ theory & Jensen--Shannon div.$^{\dagger}$  & 4 & $0.9999$ & $3.6\times10^{-4}$ / $3.7\times10^{-4}$ \\
\bottomrule
\end{tabular}
\end{table}

\subsection{Accuracy of the information-theoretic metrics}
\label{sec:acc-information}
\begin{figure}[htbp]
\centering
\includegraphics[width=0.9\textwidth]{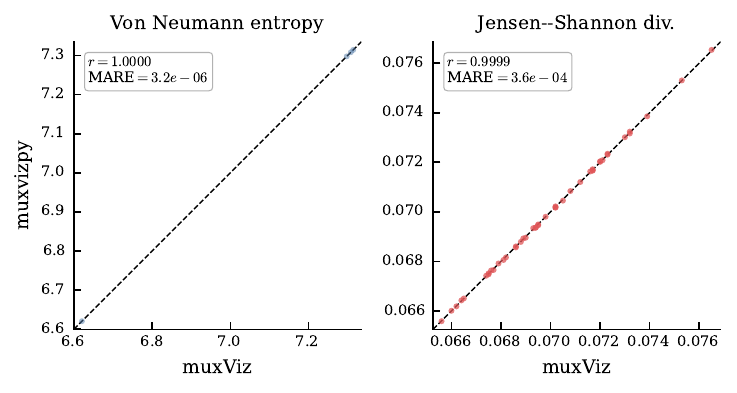}
\caption{Numerical agreement between \texttt{muxvizpy} ($y$) and \texttt{muxViz} ($x$) for the information-theoretic metrics at the largest shared network. The dashed line is the identity; the box reports Pearson $r$ and MARE.}
\label{fig:acc-information}
\end{figure}

Von Neumann entropy and Jensen--Shannon divergence are the two information-theoretic metrics computed by both tools (Table~\ref{tab:correspondence}), so unlike the CP, percolation and component metrics they do admit a direct comparison against \texttt{muxViz}. Figure~\ref{fig:acc-information} reports this agreement on the BA multiplexes (\texttt{muxViz} fails on the ER instances): the $Q=1$ von Neumann entropy reproduces the \texttt{muxViz} value to machine precision (MARE $\sim\!10^{-6}$; Section~\ref{sec:def-information}) and the layer-pairwise Jensen--Shannon divergence agrees to $\sim\!10^{-4}$ ($r=0.9999$), consistent with Table~\ref{tab:accuracy}.

\subsection{Accuracy of the approximate Katz solvers}
\label{sec:acc-katz}

\begin{figure}[htbp]
\centering
\includegraphics[width=0.9\textwidth]{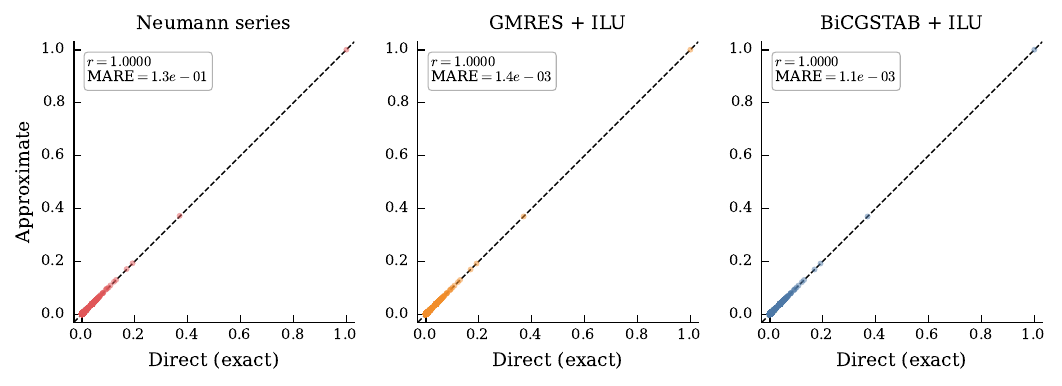}
\caption{Agreement of the approximate Katz solvers ($y$) with the exact direct solver ($x$) at the largest shared network. GMRES and BiCGSTAB lie on the identity line; the Neumann series shows value-level deviation on this large, ill-conditioned instance despite preserving rank.}
\label{fig:acc-katz}
\end{figure}

Taking the exact direct solver as reference, the preconditioned Krylov methods reproduce Katz centrality essentially exactly (GMRES: MARE median $2.0\times10^{-4}$; BiCGSTAB: $3.4\times10^{-4}$; $r=1.0000$ throughout). The Neumann series preserves the node \emph{ranking} ($r$ median $0.9999$) but its value-level error grows on the largest, most ill-conditioned instances (MARE up to $0.13$, $r$ as low as $0.81$) because a fixed iteration budget is insufficient for its slow, $\varepsilon$-limited convergence rate. Figure~\ref{fig:acc-katz} shows this contrast. In practice the preconditioned Krylov path is therefore the recommended default: it matches the exact solver while scaling like the approximate ones.

\subsection{Validation of the CP decomposition}
\label{sec:acc-decomposition}
The sparse CP/PARAFAC decomposition has no \texttt{muxViz} counterpart, so we validate it against the reference implementation in \code{tensorly} \citep{kossaifi2019tensorly}. We assemble a synthetic four-way tensor $\ten{M}\in\R^{N\times L\times N\times L}$ ($N=80$, $L=5$) from $R=4$ planted rank-one terms in the CP form of Eq.~\eqref{eq:cp}, with factor columns normalised to unit Euclidean norm, so that the exact decomposition is essentially unique. Both \texttt{muxvizpy}'s sparse ALS and \code{tensorly}'s \code{parafac} are run at rank $R$. Write the reference factor matrices as $\mat{U}^{(n)}$ and the \texttt{muxvizpy} factors as $\hat{\mat{U}}^{(n)}$ for the four modes $n=1,\dots,4$ (source/target node and source/target layer), with $\mat{u}^{(n)}_r$ and $\hat{\mat{u}}^{(n)}_s$ their $r$-th and $s$-th columns.

CP factors are identifiable only up to a common permutation of the $R$ components and, within each component, a per-mode sign and scale whose product reproduces the weight $\lambda_r$; we normalise these ambiguities away before comparing. \emph{Scale} is fixed by rescaling every factor column to unit norm, $\mat{u}^{(n)}_r\!\leftarrow\!\mat{u}^{(n)}_r/\lVert\mat{u}^{(n)}_r\rVert_2$. \emph{Permutation} is resolved through the cross-mode congruence between a reference component $r$ and a test component $s$,
\begin{equation}
  \Phi_{rs}=\prod_{n=1}^{4}
  \frac{\bigl\lvert\,\langle\mat{u}^{(n)}_r,\ \hat{\mat{u}}^{(n)}_s\rangle\,\bigr\rvert}
       {\lVert\mat{u}^{(n)}_r\rVert_2\,\lVert\hat{\mat{u}}^{(n)}_s\rVert_2},
  \label{eq:cp-congruence}
\end{equation}
the product of the per-mode cosine similarities (the Tucker congruence coefficient for the CP model). The optimal matching $\pi$ maximises the total congruence $\sum_{r}\Phi_{r\pi(r)}$ and is obtained as a linear-sum (Hungarian) \cite{hungarianassignment} assignment on the cost matrix $1-\Phi_{rs}$. \emph{Sign} is then fixed for each matched pair by choosing $\varepsilon^{(n)}_r\in\{\pm1\}$, with $\prod_{n}\varepsilon^{(n)}_r=1$, so that every per-mode inner product $\langle\mat{u}^{(n)}_r,\ \varepsilon^{(n)}_r\hat{\mat{u}}^{(n)}_{\pi(r)}\rangle$ is non-negative.

\begin{figure}[htbp]
\centering
\includegraphics[width=0.5\textwidth]{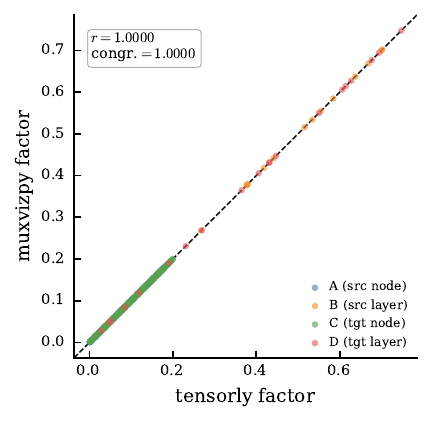}
\caption{CP factor agreement between \texttt{muxvizpy}'s sparse ALS ($y$) and \code{tensorly}'s \code{parafac} ($x$) on a planted rank-$R$ tensor, after Hungarian component alignment and sign/scale normalisation. Each point is one entry of a recovered factor column, coloured by tensor mode; the dashed line is
the identity.}
\label{fig:acc-decomposition}
\end{figure}

Agreement is then quantified at the entry level. Stacking all aligned factor entries across the four modes and $R$ components into paired samples
$\{(x_i,y_i)\}_{i=1}^{M}$---$x_i$ from \code{tensorly}, $y_i$ from \texttt{muxvizpy}---we report their Pearson correlation
\begin{equation}
  r=\frac{\sum_i (x_i-\bar{x})(y_i-\bar{y})}
         {\sqrt{\sum_i (x_i-\bar{x})^2}\,\sqrt{\sum_i (y_i-\bar{y})^2}},
  \label{eq:cp-r}
\end{equation}
together with the mean matched congruence $\bar{\Phi}=\tfrac{1}{R}\sum_{r}\Phi_{r\pi(r)}$.
The two factorisations coincide across all four modes (Figure~\ref{fig:acc-decomposition}).

\begin{remark}[Metrics without a \texttt{muxViz} accuracy reference] $k$-core, global edge overlap, local clustering, CP decomposition, vertex percolation, and the largest viable/intersection components are either absent from the standard \texttt{muxViz} API or return values not retained by the benchmark harness. They are validated differently: CP factors against \code{tensorly} (Section~\ref{sec:acc-decomposition}, Figure~\ref{fig:acc-decomposition}); the remaining metrics through unit tests on small networks with analytically known values, and, for LVC, percolation and community structure, through the biological use case of the main
text.
\end{remark}

\clearpage
%% ==================================================================
\section{Function-name correspondence (Extended Table~1)}
\label{sec:correspondence}

Table~\ref{tab:correspondence} gives the full mapping between \texttt{muxvizpy} functions and their \texttt{muxViz} equivalents, extending the abbreviated Table~1 of the main text. A dash denotes a metric with no direct \texttt{muxViz} counterpart in the standard API.

\begin{table}[ht]
\centering
\caption{Correspondence between \texttt{muxvizpy} functions and \texttt{muxViz} equivalents. ``Cost-only'' marks a metric benchmarked for computational cost but not numerical agreement.}
\label{tab:correspondence}
\footnotesize
\begin{tabular}{@{}l l l@{}}
\toprule
\textbf{muxvizpy} & \textbf{muxViz} & \textbf{Category} \\
\midrule
\code{compute\_eigenvector\_centrality}   & \code{GetMultiEigenvectorCentrality} & Versatility \\
\code{compute\_katz\_centrality}          & \code{GetMultiKatzCentrality}        & Versatility \\
\code{compute\_multipagerank\_centrality} & \code{GetMultiPageRankCentrality}    & Versatility \\
\code{compute\_multi\_rw\_centrality}     & \code{GetMultiRWCentrality} (cost-only) & Versatility \\
\code{compute\_multi\_hub\_centrality}    & \code{GetMultiHubCentrality}         & Versatility \\
\code{compute\_multi\_authority\_centrality} & \code{GetMultiAuthCentrality}     & Versatility \\
\code{get\_multi\_Kcore\_centrality}      & ---                                   & Topology \\
\code{compute\_aggregated\_multidegree}   & \code{GetMultiDegreeSum}             & Topology \\
\code{compute\_aggregated\_indegree}      & \code{GetMultiInDegreeSum}           & Topology \\
\code{compute\_aggregated\_outdegree}     & \code{GetMultiOutDegreeSum}          & Topology \\
\code{get\_modules}                       & \code{GetMultiCommunity}             & Mesoscale \\
\code{compute\_local\_clustering\_coefficient} & \code{GetLocalClustering}       & Mesoscale \\
\code{get\_LVC} / \code{get\_LIC}         & ---                                   & Topology \\
\code{compute\_average\_global\_clustering\_coefficient} & \code{GetAverageGlobalClustering} & Global \\
\code{compute\_average\_global\_overlap}  & ---                                   & Global \\
\code{compute\_vn\_entropy}               & \code{GetRenyiEntropyFromAdjacencyMatrix} & Info.\ theory \\
\code{compute\_js\_divergence}            & \code{GetJensenShannonDivergence}    & Info.\ theory \\
\code{cp\_als} (PARAFAC)                  & ---                                   & Decomposition \\
\code{vertex\_percolation}                & ---                                   & Percolation \\
\bottomrule
\end{tabular}
\end{table}

\clearpage
%% ==================================================================
\bibliographystyle{oup-abbrvnat}
\bibliography{reference}